\documentclass[prd, twocolumn, preprintnumbers, 
superscriptaddress, nofootinbib, showpacs]{revtex4}%

\usepackage{rotating}
\usepackage{lscape}

\usepackage{amssymb,hyperref, amsmath}
\usepackage{color}
\usepackage{graphicx}
\usepackage{multirow}

\begin{document}

\bibliographystyle{JHEP}

\preprint{JLAB-THY-13-1786}
\preprint{TCDMATH 13-11}

\title{Toward the excited isoscalar meson spectrum from lattice QCD}

\author{Jozef~J.~Dudek}
\email{dudek@jlab.org}
\affiliation{Jefferson Laboratory, 12000 Jefferson Avenue,  Newport News, VA 23606, USA}
\affiliation{Department of Physics, Old Dominion University, Norfolk, VA 23529, USA}

\author{Robert~G.~Edwards}
\affiliation{Jefferson Laboratory, 12000 Jefferson Avenue,  Newport News, VA 23606, USA}

\author{Peng~Guo}
\affiliation{Jefferson Laboratory, 12000 Jefferson Avenue,  Newport News, VA 23606, USA}

\author{Christopher~E.~Thomas}
\affiliation{School of Mathematics, Trinity College, Dublin 2, Ireland}

\collaboration{for the Hadron Spectrum Collaboration}

\begin{abstract}
We report on the extraction of an excited spectrum of isoscalar
mesons using lattice QCD. Calculations on several lattice volumes are performed with a range of light quark masses corresponding to pion masses down to $\sim 400$ MeV. The \emph{distillation} method enables us to
evaluate the required disconnected contributions with high statistical
precision for a large number of meson interpolating fields.
We find relatively little mixing between $\tfrac{1}{\sqrt{2}}(u\bar{u} +
d\bar{d})$ and $s\bar{s}$ in most $J^{PC}$ channels; one notable
exception is the pseudoscalar sector where the approximate $SU(3)_F$
octet, singlet structure of the $\eta,\, \eta'$ is reproduced.
We extract exotic $J^{PC}$ states, identified as hybrid mesons in which
an excited gluonic field is coupled to a color-octet $q\bar{q}$ pair,
along with non-exotic hybrid mesons embedded in a $q\bar{q}$-like spectrum.\end{abstract}



\maketitle 



\section{Introduction}

Isoscalar mesons, those which have all flavor quantum numbers equal to zero, offer a rich probe of the non-perturbative physics of QCD. Quarks of all flavors can in principle contribute, as can bound configurations constructed entirely from glue, and since quark-antiquark pairs of no net flavor can annihilate, the quark and glue sectors can mix dynamically. By studying the spectrum and hidden-flavor content of isoscalar mesons we can infer a phenomenology of annihilation dynamics within QCD. 

Experimentally there appears to be significant regularity in the excited meson spectrum, with each isovector meson of a given $J^{PC}$ typically partnered by an isoscalar meson of roughly the same mass, with both states dominantly decaying into final states not featuring strange hadrons. At a slightly larger mass, usually roughly 200 MeV higher, there is another isoscalar meson dominantly decaying into final states featuring strangeness. The most famous example of this is the $\rho, \omega, \phi$ system where the admixture of $|s\bar{s}\rangle$ into the dominantly  $\tfrac{1}{\sqrt{2}}\left( |u\bar{u}\rangle + |d\bar{d}\rangle \right)$ $\omega$ meson is estimated to be less than $1\%$. Other reasonably well-determined examples include the tensor mesons $a_2(1320), \, f_2(1270),\, f'_2(1525)$, and the $\rho_3(1690),\, \omega_3(1670),\, \phi_3(1850)$ system \cite{PDG}.

There are only a few known exceptions to this pattern of ``ideal mass mixing". The low-lying pseudoscalar sector, $\pi,\,\eta,\,\eta'$, is the best studied, with the conventional interpretation being that the $\eta$ and $\eta'$ are mixed such that they are close to the octet $\tfrac{1}{\sqrt{6}} \left( |u\bar{u}\rangle + |d\bar{d}\rangle - 2 |s\bar{s}\rangle \right)$ and singlet  $\tfrac{1}{\sqrt{3}} \left( |u\bar{u}\rangle + |d\bar{d}\rangle + |s\bar{s}\rangle \right)$ representations of $SU(3)_F$ respectively. The lightest isoscalar $1^{++}$ mesons, $f_1(1285),\, f_1(1420)$, appear to be mixed somewhere between ideal in mass and ideal in $SU(3)_F$ -- their relative rates of decay into $\gamma \rho$ and $\gamma \phi$ suggest there may be as much as $\sim 15\%$ hidden strange in the lighter state \cite{Close:1997nm,PDG}.

The scalar sector ($0^{++}$) does not admit a simple description -- while some features are now largely agreed upon, such as the existence of a broad $\sigma$ resonance (see e.g. \cite{Caprini:2005zr}) and narrow $f_0(980), a_0(980)$ resonances near the $K\overline{K}$ threshold, the exact number of scalar resonances between 1 and 2 GeV, and their flavor composition, is not a settled question \cite{PDG, Amsler:2004ps, Bugg:2004xu, Close:2002zu, Klempt:2007cp, Crede:2008vw}.

The level of theoretical understanding of isoscalar mesons is generally much lower than for the isovector mesons. The gross features of the excited isovector spectrum can be reasonably well described by proposing that there exist effective heavy ``constituent quark" degrees-of-freedom, from which we can build mesons as quark-antiquark pairs with relative orbital angular momentum. Ideal mass mixed isoscalar mesons, $\tfrac{1}{\sqrt{2}}(u\bar{u}+d\bar{d})$, $s\bar{s}$, can then be described if we state, by fiat, that annihilation contributions are negligible, ensuring no flavor mixing and no splitting between the lighter isoscalar and the isovector state. The hidden-strange states are then heavier solely by virtue of the strange constituent quark being heavier. With this approach, there is no natural way to explain the exceptions to ideal mass mixing, such as the $1^{++}$ sector, or the non-trivial structure of the scalar sector.
 
The constituent quark picture fails to describe the low-lying pseudoscalar sector, but here we have a solid phenomenology based upon them being pseudo-Goldstone bosons of spontaneously broken chiral symmetry. The approximate $SU(3)$ axial-vector symmetry that is broken by the vacuum gives rise to an octet of light pseudoscalar mesons that includes the pions, kaons and the $\eta$. The remaining approximate $U(1)$ axial-vector symmetry, whose breaking we might expect to give us a light singlet pseudoscalar, is in fact not a symmetry at the quantum level, being broken by an anomalous divergence which receives contributions from topologically non-trivial gauge-field configurations \cite{'tHooft:1976up, 'tHooft:1986nc}. This provides a mechanism by which the dominantly singlet  $\eta'$ can be heavier than the dominantly octet $\eta$ as is observed experimentally.

States of pure glue, ``glueballs", are well studied using lattice methods applied to the quark-less theory of $SU(3)$ color Yang-Mills (e.g. \cite{Morningstar:1999rf}). A clear spectrum of color-singlet bound states is predicted, with the lightest state being a scalar which may have a mass under 2~GeV. Other relatively low-lying states include a tensor and a pseudoscalar, before the spectrum becomes dense at high energies. Experimentally it is not clear that there is a single scalar resonance that can be identified with a glueball, and it remains a longstanding problem to find a clear observable filter that marks out a state as having a glueball structure rather than $q\bar{q}$ \cite{Crede:2008vw}. Determining how glueballs enter into the spectrum in full QCD with dynamical quark fields demands a complete calculation extracting the entire low-energy spectrum of color-singlet states, something which has not been seriously attempted to date.

In short, outside of the low-energy pseudoscalar sector, where (spontaneously broken) symmetries of QCD constrain the possibilities, little is known about isoscalar mesons from first principles in QCD.

Lattice QCD enables us to determine the excited spectrum of QCD in a controlled approximation. Excited state energies are computed from the exponential decay of Euclidean time correlators featuring composite QCD operators with hadron quantum numbers. The gauge coupling (and hence the lattice spacing), lattice volume, and the quark masses are inputs to the calculations that can be systematically varied to approach the limit in which physical QCD is duplicated. In this limit, direct comparison with experiment can be made.

Recently significant progress has been made in approaching this limit in determinations of the lightest baryons and isovector mesons (e.g. \cite{Durr:2008zz}). Away from this limit, with light quarks somewhat heavier than physical, detailed spectra of {\it excited} isovector mesons and other excited non-isoscalar hadrons have been obtained \cite{ Dudek:2009qf,  Dudek:2010wm, Dudek:2012ag, Thomas:2011rh, Edwards:2011jj, Edwards:2012fx, Engel:2013ig, Engel:2011aa}. Extracting the {\it isoscalar} meson spectrum remains a challenging prospect -- distinguishing isoscalar from isovector correlators requires the evaluation of disconnected Wick contractions, and it has proven difficult to obtain signals of high statistical precision for these (but see recent progress, restricted to the lightest pseudoscalars, in \cite{Christ:2010dd, Ottnad:2012fv,Gregory:2011sg}). Furthermore in order to reliably extract a spectrum of {\it excited} isoscalar states, one should compute a large number of isoscalar correlators using a range of quark-gluon composite operators to interpolate states. In this case we demand a method that provides a computationally efficient means to compute quark propagation from arbitrarily complicated operators. In this work we will take advantage of a number of useful features of the {\it distillation} quark smearing framework \cite{Peardon:2009gh} to address the difficulties posed above. We find that it is possible to obtain an excited state isoscalar spectrum of high statistical quality with a level of computational effort that can be satisfied using the capability provided by Graphics Processing Units. 

In this paper, building on our initial exploratory efforts in \cite{Dudek:2011tt}, we will present isoscalar meson spectra obtained for a range of light quark masses on several lattice volumes. We explore systematics observed in the spectrum, compare with the isovector spectrum and extract information about the hidden light/strange composition of isoscalar mesons. 

Among the interesting features present in the isovector meson spectrum reported on in \cite{Dudek:2009qf,Dudek:2010wm,Dudek:2011bn} are a set of exotic $J^{PC}$ states; that is, states whose quantum numbers cannot be described by a quark-antiquark pair with orbital angular momentum, and which thus lie outside the otherwise rather successful constituent-quark picture. These additional states were found to have strong overlap onto operators featuring a chromomagnetic gluonic construction coupled to the quark fields, as were a number of otherwise unexpected states with non-exotic $J^{PC}$. An explanation was proposed that these states are \mbox{{\it hybrid}} mesons, in which a quark-antiquark pair is coupled to an excited gluonic field, a sector of QCD long expected to exist \cite{Horn:1977rq,Isgur:1984bm,Barnes:1982tx,Chanowitz:1982qj,General:2006ed,Guo:2008yz}. Experimentally there are hints that at least one such state with $1^{-+}$ may be present (for a critical review see \cite{Meyer:2010ku}). It is important to also determine if isoscalar partners of these isovector hybrid mesons exist, and in what mass region -- in this paper we will report on a clear identification of such states. 

Because of the complexity of the scalar ($0^{++}$) sector, which experimentally contains a low-lying broad resonance and where a low-lying glueball is
expected, we defer a detailed investigation of scalar mesons to future work.

The remainder of the manuscript is arranged as follows: Section \ref{lat} describes how one computes isoscalar correlation functions in lattice QCD, describes the distillation construction, presents some selected results showing the signal quality obtained in explicit computation, and discusses correlation functions featuring glueball-like operators. Section \ref{spec} describes the extraction of the excited state spectrum from a matrix of correlation functions and the determination of the hidden flavor mixing in isoscalar states. The method is demonstrated with results from a single symmetry sector before a discussion of the caveats which should be applied to an interpretation of the extracted spectrum. Section \ref{results} presents the spectra determined for a range of quark masses and lattice volumes. Section \ref{pheno} contains a phenomenological description of the observed spectra in terms of constituent $q\bar{q}$ constructions supplemented with hybrid mesons. Finally, in Section \ref{outlook}, we conclude and present possible directions for future calculations to address the physics of isoscalar mesons.

\pagebreak
\section{Lattice technology} \label{lat}

\begin{figure}[h]
 \centering
\includegraphics[width=.27\textwidth
]{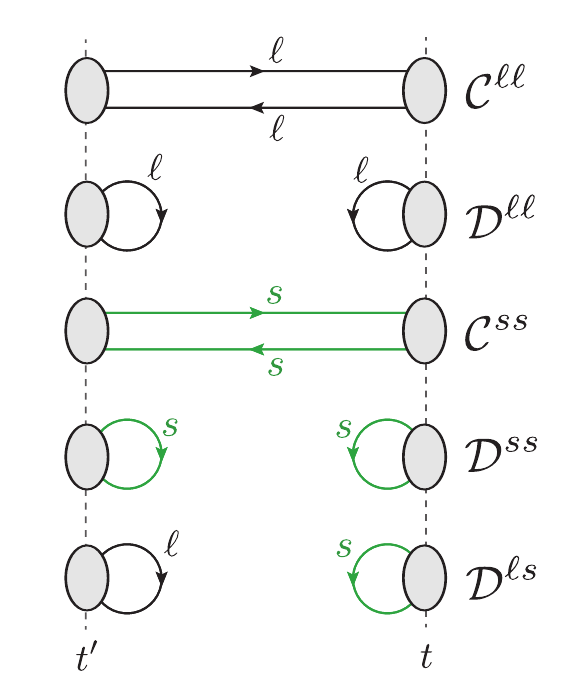} 
\caption{Quark propagation for correlation functions built using the operator basis in Eqn \ref{massbasis}. Each line represents a quark propagator for either a light (black) or a strange (green) quark.   \label{wicks}}
\end{figure}

The basic object we will use to extract the spectrum of isoscalar meson eigenstates of QCD is a Euclidean correlation function, 
\begin{equation}
	C_{ij}(t',t) = \big\langle 0 \big| \mathcal{O}_i(t') \mathcal{O}^\dag_j(t) \big| 0 \big\rangle , \label{corr}
\end{equation}
where the interpolating fields, $\mathcal{O}_i^\dag$, are gauge-invariant combinations of the basic quark and gluon fields of QCD, constructed to transform with $I=0$ under rotations in quark flavor space. The complete set of discrete eigenstates, $|\mathfrak{n}\rangle$, with the quantum numbers of $\mathcal{O}_{i,j}$ appears in the spectral decomposition,
\begin{equation}
C_{ij}(t',t) = \sum_\mathfrak{n} \frac{1}{2 E_\mathfrak{n}} \big\langle 0 \big| \mathcal{O}_i(0) \big| \mathfrak{n} \big\rangle \big\langle \mathfrak{n} \big| \mathcal{O}^\dag_j(0) \big| 0 \big\rangle\,  e^{-E_\mathfrak{n} (t'-t)}. \label{spectraldecomp}
\end{equation}
The simplest isoscalar meson operators are of quark-bilinear structure, $\bar{\psi} \mathbf{\Gamma} \psi$, and with two light-quark flavors, $u,d$ (assumed here to be degenerate, so we have an exact isospin symmetry), and one heavier flavor, $s$, a possible flavor basis having $I=0$ is given by
\begin{equation}
 \mathcal{O}^\ell = \tfrac{1}{\sqrt{2}}\big( \bar{u} \mathbf{\Gamma} u + \bar{d} \mathbf{\Gamma} d \big),\; \mathcal{O}^s = \bar{s} \mathbf{\Gamma} s . \label{massbasis}
\end{equation}
After functional integration of the quark fields, correlation functions can be expressed in terms of {\it connected} components, $\mathcal{C}$, diagonal in quark flavor, and {\it disconnected} components, $\mathcal{D}$, which can mix flavor. The Wick contractions corresponding to $\mathcal{C}, \mathcal{D}$ are shown schematically in Figure \ref{wicks}.

The four possible correlation functions using the flavor basis in Eqn \ref{massbasis}, $C^{q'q}(t',t) = \big\langle 0 \big| \mathcal{O}^{q'}(t') \mathcal{O}^{q\dag}(t) \big| 0 \big\rangle$, with $q=\ell,s$ can be expressed as a matrix,
\begin{equation}
\begin{pmatrix} C^{\ell \ell} & C^{\ell s} \\ C^{s \ell} & C^{s s}  \end{pmatrix} = \begin{pmatrix} - \mathcal{C}^{\ell \ell} + 2 \mathcal{D}^{\ell \ell} & \sqrt{2} \, \mathcal{D}^{\ell s} \\ \sqrt{2}\,  \mathcal{D}^{s \ell} &  - \mathcal{C}^{s s} + \mathcal{D}^{s s}\end{pmatrix}. \label{flavormatrix}
\end{equation}
The off-diagonal elements of this matrix allow the eigenstates of the theory to be admixtures of light and strange quarks, with the degree of mixing being determined dynamically. Note that the connected contribution, $-\mathcal{C}^{\ell \ell}$, is precisely the contraction required to compute the {\it isovector} correlation function corresponding to any of the flavor constructions, {$\bar{u} \mathbf{\Gamma} d,\, \bar{d} \mathbf{\Gamma} u, \tfrac{1}{\sqrt{2}}\big( \bar{u} \mathbf{\Gamma} u - \bar{d} \mathbf{\Gamma} d \big)$}, and thus, even without strange quarks, the inclusion of the disconnected contribution, $\mathcal{D}^{\ell \ell} $, means the isoscalar spectrum need not be the same as the isovector spectrum.

An alternative basis, corresponding to a simple orthogonal transformation of Eqn \ref{massbasis}, is
\begin{align}
 \mathcal{O}^\mathbf{1} &= \tfrac{1}{\sqrt{3}}\big( \bar{u} \mathbf{\Gamma} u + \bar{d} \mathbf{\Gamma} d  + \bar{s} \mathbf{\Gamma} s \big) ,  \nonumber\\
 \mathcal{O}^\mathbf{8} &=\tfrac{1}{\sqrt{6}}\big( \bar{u} \mathbf{\Gamma} u + \bar{d} \mathbf{\Gamma} d  - 2 \bar{s} \mathbf{\Gamma} s \big) .  \label{SU3basis}
\end{align}
where the labels $\mathbf{1}, \mathbf{8}$ allude to the fact that these combinations are singlet and octet irreducible representations of an $SU(3)$ flavor symmetry. This basis has the property that in the limit that the $u,d$ and $s$ quarks all have the same mass, only the correlator $C^{\mathbf{1},\mathbf{1}}$ has a disconnected contribution and $C^{\mathbf{1},\mathbf{8}}$ is zero.

We compute correlation functions using {\it distillation} \cite{Peardon:2009gh} -- within this framework the quark fields on a timeslice entering in the interpolating functions are smeared over space by an operator, $\Box$, whose purpose is to enhance the low momentum quark and gluon modes that dominate low mass hadrons. The operator, $\Box$, acting in color and position space, is constructed as 
\begin{equation}
\Box(t)    = \sum_{k=1}^N \xi^{k}(t) \, \xi^{k\dag}(t), \label{Box}
\end{equation}
where  $ \xi^{k}(t)$ are the lowest $N$ eigenvectors of the gauge-covariant laplacian evaluated on the background of the spatial gauge-fields of timeslice $t$. Quark bilinear operators can be constructed in distillation as
\begin{equation}
\bar{\psi} \Box \mathbf{\Gamma} \Box \psi,	
\end{equation}
and the outer-product definition of $\Box$, Eqn \ref{Box}, allows a factorization of the correlation functions into products of objects describing quark propagation and objects encoding the particular operator constructions used. The propagation objects, known as {\it perambulators},
\begin{equation}
	\tau_{\alpha\beta}^{ij}(t',t) = \xi^{i\dag}(t')\, \big[M^{-1}\big]_{\alpha\beta}(t',t)\, \xi^{j}(t),
\end{equation}
are obtained by inverting the Dirac matrix, $M$, whose Dirac spin indices are here exposed, on a finite number of sources, $\xi^{j}(t)$. The connected and disconnected contributions to isoscalar correlation functions then take the form
\begin{align}
\mathcal{C}^{q' q}_{AB}(t',t) &= \delta_{q q'} \,\mathrm{Tr}\big[  \Phi_A(t')\, \tau^{q}(t',t)\, \Phi_B(t)\, \tau^q(t,t') \big] \nonumber \\
\mathcal{D}^{q' q}_{AB}(t',t) &=  \mathrm{Tr}\big[  \Phi_A(t')\, \tau^{q'}(t',t') \big] \, \mathrm{Tr}\big[\Phi_B(t)\, \tau^q(t,t) \big] , \label{conndisc}
\end{align}
where contracted indices have been suppressed and where the choice of operator construction is encoded in $(\Phi_A)^{ij}_{\alpha\beta}(t) = \xi^{i\dag}(t)\, \mathbf{\Gamma}^A_{\alpha\beta}(t) \, \xi^{j}(t)$. There is considerable flexibility allowed in the choice of $\mathbf{\Gamma}$ -- in this work we will use some number of gauge-covariant derivatives and a projection over position-space into zero momentum. To the extent that the construction, $\mathbf{\Gamma}$, is well supported in the finite basis of distillation vectors, $\{\xi^{i=1\ldots N}\}$,  distillation is a very efficient means to compute the correlation functions above since the traces are over the set of $N$ distillation vectors which is much smaller than the full lattice space.

We can obtain a determination of the connected contribution, $\mathcal{C}(t, t_0)$, for all values of $t$, using only perambulators from a single time-source 
  at $t_0$. On the other hand, in order to study the $t$-dependence of $\mathcal{D}(t, t_0)$, Eqn~\ref{conndisc} indicates that we require perambulators from sources at every value of $t$. Operationally, since we compute all such perambulators, we opt to maximize signal over noise by averaging over many time sources:
\begin{align}
\mathcal{C}^{q q}(t) &\equiv \frac{1}{N_\mathcal{C}} \sum_{ \{t_0\}}^{N_\mathcal{C}} \mathcal{C}^{q q}(t + t_0, t_0)  \nonumber \\
\mathcal{D}^{q' q}(t) &\equiv \frac{1}{N_t} \sum_{ \{t_0\}}^{N_t} \mathcal{D}^{q' q}(t + t_0, t_0). \nonumber
\end{align}
The disconnected piece is averaged over all $N_t = 128$ time sources, while the connected piece needs far less averaging to achieve a comparable level of statistical fluctuation -- the number of equally spaced sources used, $N_\mathcal{C}$, is given in Table \ref{tab:lattices}.

\begin{table}[t]
  \begin{tabular}{cccc|cccc}
    $\substack{a_t m_\ell\\a_t m_s}$ & $\substack{m_\pi\\/\mathrm{MeV}}$ & $\frac{m_K}{m_\pi}$ &  $(L_s/a_s)^3 \times (L_t/a_t)$   &$N_{\mathrm{cfgs}}$ & $N_\mathcal{C}$ & $N_{\mathrm{vecs}}$ \\
    \hline\hline
	
\multirow{2}{*}{ $\substack{-0.0743\\-0.0743}$}  & \multirow{2}{*}{702} & \multirow{2}{*}{1}  	& $16^3\times 128$ & 535 & 32 & 64 \\
&&														      	& $20^3 \times 128$ & 505 & 8 & 128 \\
   
     \hline	
     $\substack{-0.0808\\-0.0743}$ & 524 & 1.15  & $16^3\times 128$ & 500 & 32 & 64 \\			
     \hline
     
     \multirow{3}{*}{ $\substack{-0.0840\\-0.0743}$}  & \multirow{3}{*}{391} & \multirow{3}{*}{1.40}  	& $16^3\times 128$ & 479 & 32 & 64 \\
&&														      	& $20^3 \times 128$ & 600 & 8 & 128 \\
&&														      	& $24^3 \times 128$ & 553 & 8 & 162 
    
  \end{tabular}
  \caption{Lattice gauge field parameters. Also shown the number of time sources averaged over, $N_\mathcal{C}$, in the computation of connected correlators, $\mathcal{C}$, and the number of distillation vectors, $N_\mathrm{vecs}$}
\label{tab:lattices}
\end{table}

\begin{table}
\begin{tabular}{r|cccccccc}
& $a_0$ & $\pi$ & $\pi_2$ & $b_0$ & $\rho$ & $\rho_2$ & $a_1$	& $b_1$\\ 
\hline
$\Gamma$ & $1$ &$\gamma_5$ & $\gamma_0\gamma_5$ & $\gamma_0$ &$\gamma_i$& $\gamma_0 \gamma_i$ & $\gamma_5 \gamma_i$ & $\gamma_0\gamma_5 \gamma_i$
\end{tabular}
\caption{Gamma matrix naming scheme. \label{table:gamma}}
\end{table}

A powerful way to extract the spectrum of eigenstates in Eqn \ref{spectraldecomp} is to utilize a {\it matrix} of correlation functions, constructed from a basis of operators $\{ \mathcal{O}_{i=1 \ldots N_\mathrm{ops} } \}$ all having the same conserved quantum numbers. The matrix can be ``diagonalized" using a variational approach 
\cite{Michael:1985ne,Luscher:1990ck,Blossier:2009kd} to be described in the next section. 
Our choice of operator basis is described in detail in \cite{Dudek:2009qf,Dudek:2010wm} -- the fermion bilinear operators are of structure,
\begin{equation}
\bar{\psi} \Box \, \Gamma \overleftrightarrow{D} \ldots \overleftrightarrow{D} \, \Box \psi,
\label{eq:bilinear}
\end{equation}
with up to three gauge-covariant derivatives. The Dirac gamma matrix structures are listed in Table~\ref{table:gamma}. The Cartesian vectorlike gamma matrices and derivatives ${\overleftrightarrow{D} = \overleftarrow{D} - \overrightarrow{D}}$ are expressed in a circular basis so that they transform as spin $J=1$. Using standard $SO(3)$ Clebsch-Gordan coefficients, the Dirac gamma matrix structures are coupled to the derivative structures to produce operators, $\mathcal{O}^{J,M}$, where $M$ is the $z$ component of a spin-$J$ operator. The choice of $\Gamma$ and the derivative structure determine the parity, $P$, and charge conjugation, $C$, to give an operator with overall continuum quantum numbers $J^{PC}$ at zero momentum. The notation used follows that of Ref.~\cite{Dudek:2010wm}: an operator $\big(\Gamma\times D^{[N]}_{J_D}\big)^J$ contains a gamma matrix, $\Gamma$, using the naming scheme defined in Table~\ref{table:gamma}, and $N$ derivatives coupled to spin $J_D$, and overall coupled to spin $J$.

Reflecting the reduced cubic symmetry of the lattice, the continuum operators $\mathcal{O}^{J,M}$ are {\it subduced} into five irreducible representations (irreps): $A_1, T_1, T_2, E $ and $A_2$ \cite{Johnson:1982yq,Dudek:2010wm}, 
\begin{equation}
\mathcal{O}^{[J]}_{\Lambda,\lambda} =\sum\nolimits_M \mathcal{S}^{\Lambda,\lambda}_{J,M} \, \mathcal{O}^{J,M},
\end{equation}
to form operators that transform in the $\lambda^\mathrm{th}$ row of a lattice irrep, $\Lambda$. The $\mathcal{S}^{\Lambda,\lambda}_{J,M}$ are subduction matrices that are presented in Appendix A of Ref.~\cite{Dudek:2010wm}. We average the final correlation functions over each row, $\lambda$, of the irrep, $\Lambda$.

Finally, for each choice of $\mathbf{\Gamma}\equiv \big(\Gamma\times D^{[N]}_{J_D}\big)^J$ we construct two operators according to the flavor basis defined in Eqn~\ref{massbasis}.

In this work, we evaluate correlation functions on a set of dynamical anisotropic lattices which have a finer spacing in the temporal direction than the spatial directions. Improved gauge and fermion actions are used, with two mass-degenerate light dynamical quarks and one strange dynamical quark, of masses $m_\ell$ and $m_s$ respectively. Details of the formulation of the actions can be found in Refs.~\cite{Edwards:2008ja,Lin:2008pr}. The lattices have a spatial lattice spacing $a_s\sim 0.12~{\rm fm}$ with a temporal lattice spacing approximately $3.5$ times smaller, corresponding to a temporal scale $a_t^{-1}\sim 5.6$ GeV. The particular lattices used are presented in Table \ref{tab:lattices} along with parameters relevant for the correlator construction. In the remainder of the paper we will quote determined hadron mass values via a ratio to the $\Omega$ baryon mass as determined on the same lattice, scaled by the physical $\Omega$ baryon mass: $\frac{a_t m_H}{a_t m_\Omega} \cdot m_\Omega^{\mathrm{phys.}}$.

Observables are computed on gauge-field configurations separated by 20 trajectories after 1000 trajectories have been used for thermalization. It is found that the integrated autocorrelation time is small for the observables investigated in this paper. After binning the correlation functions over 10 separate measurements, no discernible change is found in the subsequent analysis of the spectrum.

\subsection{Example correlation functions}

\begin{figure}
 \centering
\includegraphics[width=.42\textwidth
]{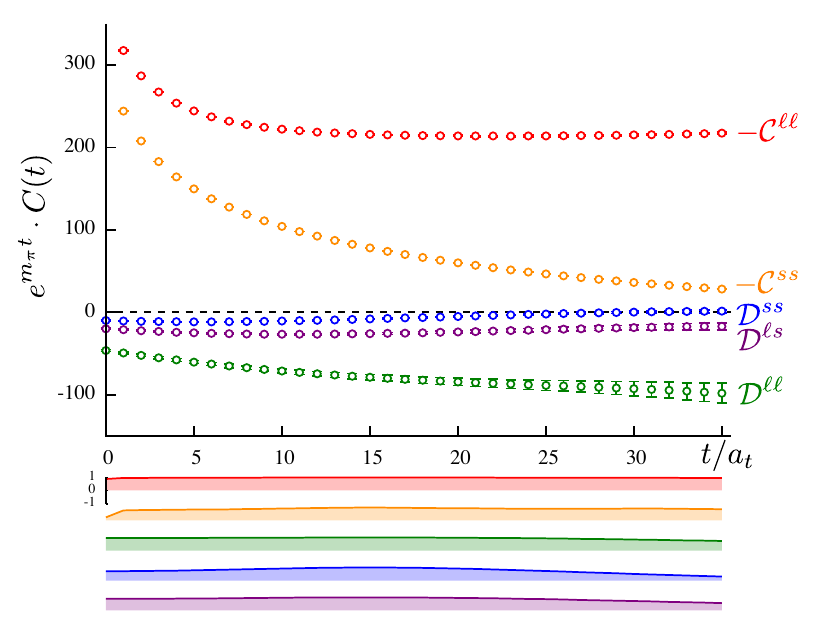} 
\caption{Connected and disconnected contributions to $A_1^{-+}$ correlation functions in Eqn \ref{flavormatrix} for the operator $\mathbf{\Gamma} = \gamma_5$. Stacked graphs under the main plot shows the timeslice correlation with a reference timeslice $t/a_t = 15$.   \label{gamma5}}
\end{figure}

\begin{figure}
 \centering
\includegraphics[width=.42\textwidth
]{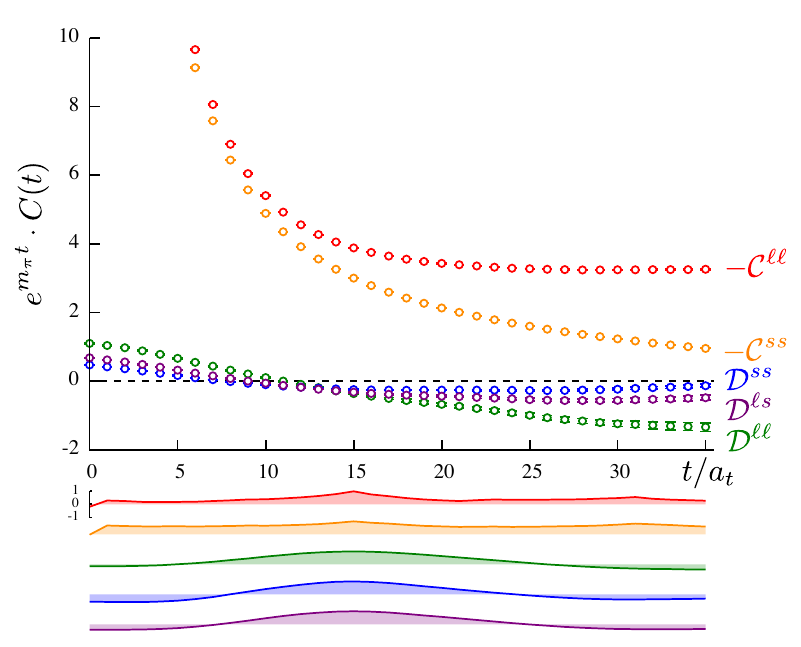} 
\caption{As Figure \ref{gamma5}, for the operator $\mathbf{\Gamma} = \big(b_1\times D^{[1]}_{J=1}\big)^{J=0}_{A_1}$ in the irrep $A_1^{-+}$. \label{A1mp_b1xD}}
\end{figure}

\begin{figure}
 \centering
\includegraphics[width=.42\textwidth
]{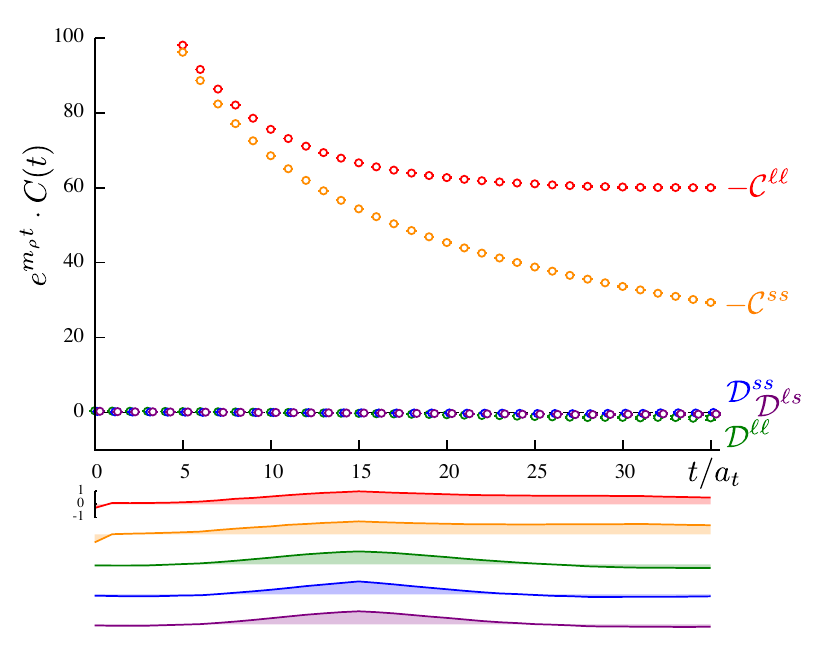} 
\caption{As Figure \ref{gamma5}, for the operator $\mathbf{\Gamma} = \gamma_i$ in the irrep $T_1^{--}$. \label{gammai}}
\end{figure}

\begin{figure}
 \centering
\includegraphics[width=.42\textwidth
]{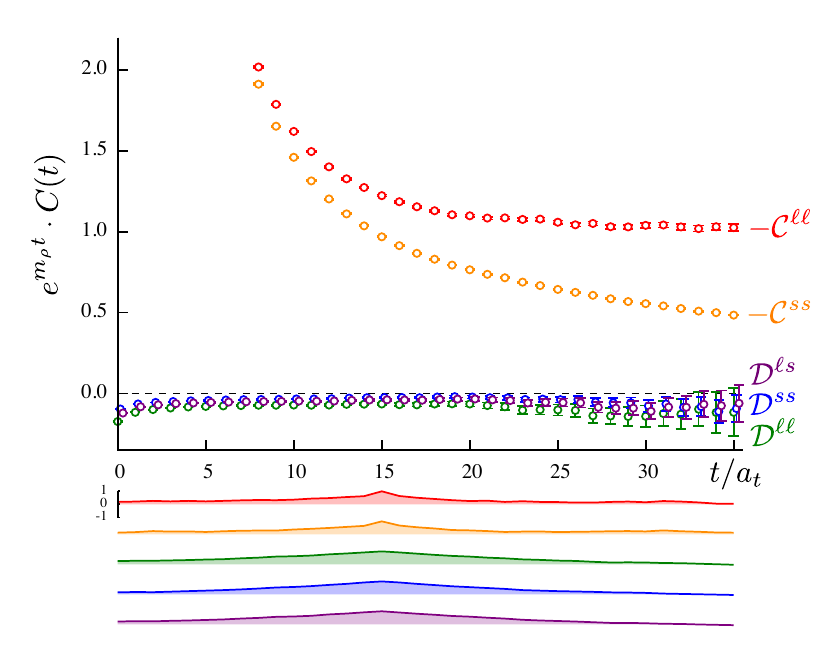} 
\caption{As Figure \ref{gamma5}, for the operator $\mathbf{\Gamma} = \big(\pi\times D^{[2]}_{J=1}\big)^{J=1}_{T_1} $ in the irrep $T_1^{--}$.   \label{T1mmhyb}}
\end{figure}

\begin{figure}
 \centering
\includegraphics[width=.42\textwidth
]{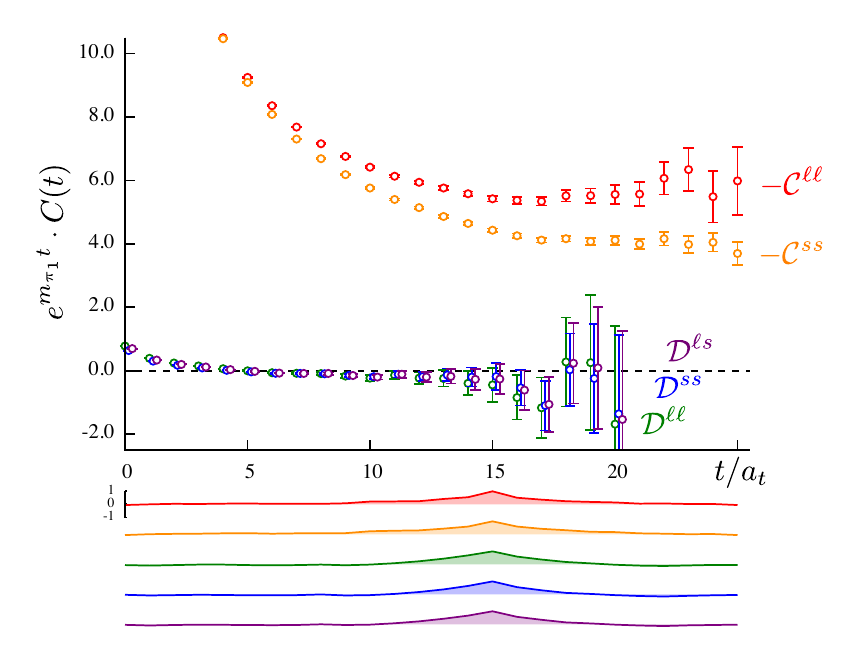} 
\caption{As Figure \ref{gamma5}, for the operator $\mathbf{\Gamma} = \big(\rho\times D^{[2]}_{J=1}\big)^{J=1}_{T_1} $ in the irrep $T_1^{-+}$.  \label{T1mphyb}}
\end{figure}

The connected and disconnected components of a set of illustrative correlation functions are shown in Figures \ref{gamma5}-\ref{T1mphyb}, computed on the $24^3\times 128$ lattice with $m_\pi = 391 \,\mathrm{MeV}$. The plots show the time-dependence, weighted by $e^{m_X t}$, where $m_X$ is the mass of the lightest isovector meson with the appropriate quantum numbers (as determined through variational analysis of the connected light-quark correlators). Also shown stacked underneath the main plot are the timeslice correlations\footnote{We compute the correlation, $\hat{\mathbb{C}}(t,t')$, via the data covariance,
\begin{equation}
\mathbb{C}(t,t') = \frac{1}{N_\mathrm{cfgs}(N_\mathrm{cfgs}-1)} \sum_{i=1}^{N_\mathrm{cfgs} } \big(C_i(t) - \overline{C(t)} \big) \big(C_i(t') - \overline{C(t')} \big), \nonumber
\end{equation}
as 
\begin{equation}
\hat{\mathbb{C}}(t,t') = \frac{ \mathbb{C}(t,t') }{\sqrt{\mathbb{C}(t,t)} \sqrt{\mathbb{C}(t',t')} } . \nonumber
\end{equation}¥
} in the data relative to a reference timeslice, $t/a_t = 15$. Data with no timeslice correlations would have this variable being $1$ at $t/a_t = 15$ and zero elsewhere.

Figure \ref{gamma5} shows correlator components for the simplest $A_1^{-+}$ operator, $\mathbf{\Gamma} = \gamma_5$. We observe that the disconnected contributions are significant, particularly $\mathcal{D}^{\ell \ell}$ and the off-diagonal flavor term, $\mathcal{D}^{\ell s}$. Compare this with Figure \ref{gammai}, which shows $\mathbf{\Gamma}=\gamma_i$, where we observe the disconnected contributions to be orders of magnitude smaller than the connected. At large times we expect the isoscalar $A_1^{-+}$ correlators to be dominated by the $\eta$ and $\eta'$ mesons which empirically are significantly heavier than their isovector cousin, the pion, and strongly admixed in the ${\tfrac{1}{\sqrt{2}} \Big( \big|u\bar{u} \big\rangle +  \big|d\bar{d} \big\rangle\Big), \big|s\bar{s}\big\rangle}$ basis. Conversely the large time behavior of the isoscalar $T_1^{--}$ correlators will be governed by the $\omega$ and $\phi$ mesons which are believed to be almost pure $\tfrac{1}{\sqrt{2}} \Big( \big|u\bar{u} \big\rangle +  \big|d\bar{d} \big\rangle\Big), \big|s\bar{s}\big\rangle$, respectively and where the $\omega$ is very similar in mass to the $\rho$. The behavior shown in Figures \ref{gamma5}, \ref{gammai} is seen to be qualitatively in agreement with these expectations. More quantitative statements will follow from variational analysis of a matrix of correlation functions and will be presented in the next section.

In Figure \ref{A1mp_b1xD} we show the set of correlation function components for another $A_1^{-+}$ operator, $\big(b_1\times D^{[1]}_{J=1}\big)^{J=0}_{A_1}$. We see that the connected component again relaxes to a mass scale compatible with $m_\pi$ at large times, as one would expect, and that at large times the relative size of the various connected and disconnected components is similar to that observed in Figure \ref{gamma5}, indicating that the same low-lying mesons are dominating the correlation functions. However, we can clearly see that the earlier time behavior is quite different to that observed in Figure \ref{gamma5}, owing to the different relative weighting of excited states in Eqn \ref{spectraldecomp} for this operator compared with $\gamma_5$. It is this variation with differing operators that we will take advantage of to determine excited state contributions in variational analysis.

In Figure \ref{gamma5} the data is not seen to fluctuate timeslice-to-timeslice to the degree one would expect on the basis of the errorbars. This suggests that the data may be correlated in time, and indeed explicit evaluation of the correlation shows this to be the case (see the stacked graphs beneath the main plot). All fitting of time-dependences in this paper will be done using correlated $\chi^2$ fits\footnote{ 
By correlated fits to $C(t)$, we mean those which use the inverse data covariance, $\mathbb{C}^{-1}$,
\begin{equation}
\chi^2\big(\{a_i\}\big) = \sum\nolimits_{t,t'} \big[ C(t) - f(t; \{a_i\}) \big] \mathbb{C}^{-1}(t,t') \big[ C(t') - f(t'; \{a_i\}) \big] \nonumber
\end{equation}
}
. Note that the high degree of correlation is not a general feature of pseudoscalar operators -- the correlators for $\mathbf{\Gamma} = \big(b_1\times D^{[1]}_{J=1}\big)^{J=0}_{A_1}$ plotted in Figure \ref{A1mp_b1xD} show much less correlation.

In previous publications \cite{Dudek:2010wm,Dudek:2011bn} we have explored the role that {\it hybrids} play in the meson spectrum. We found that non-exotic $J^{PC}$ hybrids embedded in the conventional meson spectrum can be identified by their large overlap onto operators featuring a chromomagnetic construction (e.g. $D^{[2]}_{J=1} $). In order to explore this phenomenology in the isoscalar sector we need good determinations of correlators featuring such operators. In Figure~\ref{T1mmhyb} we show connected and disconnected contributions for the $T_1^{--}$ operator $\mathbf{\Gamma} = \big( \pi \times D^{[2]}_{J=1} \big)^{J=1}$. We observe signals with rather high statistical quality, where, as one would expect, the connected term does eventually relax to the ground-state $\rho$. Note that the disconnected contributions, though small, are non-zero at intermediate times which may suggest that excited isoscalar $1^{--}$ states having good overlap onto this operator might differ somewhat from the corresponding isovector states.

Finally we examine the form of exotic quantum numbered correlators, which we expect to contain exotic hybrid mesons, with an example being $T_1^{-+}$ through the operator $\mathbf{\Gamma} = \big( \rho \times D^{[2]}_{J=1} \big)^{J=1}$. While the signals are observed to be somewhat noisier than the previous cases, there is clearly enough data (which we note is not strongly timeslice correlated) to attempt a spectrum extraction.

\subsection{``Glueball" operators}

It is possible to construct isoscalar operators within QCD without explicitly including quark fields -- suitable combinations of gauge-link fields can be built which transform irreducibly, and we might assume that such constructions would have strong overlap onto states in the spectrum which have ``glueball" structure. Unlike in the quark-less $SU(3)$ color Yang-Mills studies (e.g. \cite{Morningstar:1999rf}), we cannot study the spectrum of glueballs in isolation since there are generically $q\bar{q}$ isoscalar constructions which have the same quantum numbers and which can mix with the glueballs to give the physical eigenstates. As such the appropriate approach is to supplement the fermion bilinear operators described earlier with a set of glueball operators and explore cross-correlators.

\begin{figure}[t]
 \centering
\includegraphics[width=.49\textwidth
]{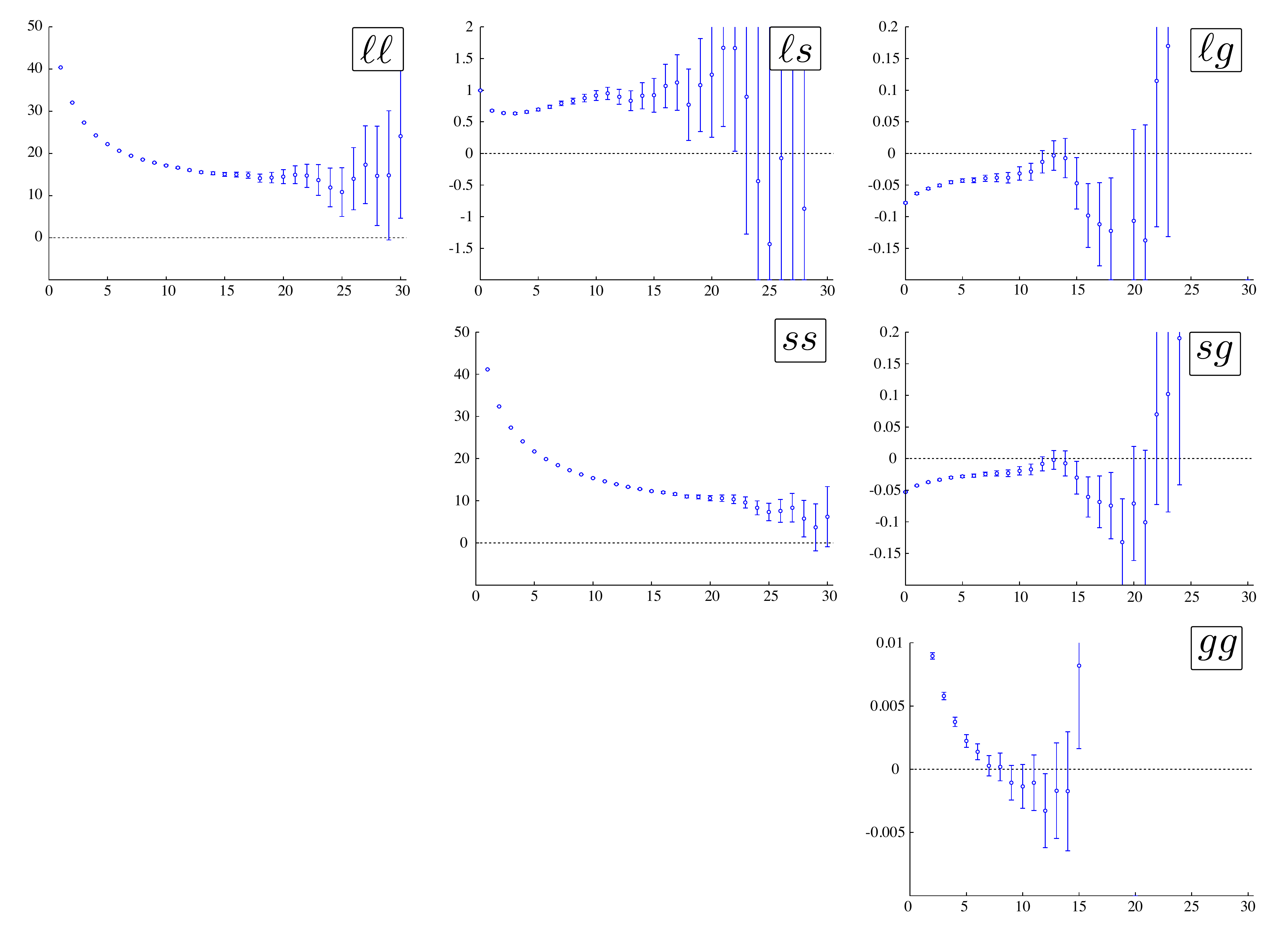} 
\caption{A selection of $E^{++}$ correlation functions with operators $\mathbf{\Gamma} = \big(\rho \times D^{[1]}_{J=1}\big)^{J=2}$ for both light ($\ell$) and strange ($s$) quark flavors, and the glueball operator $D^{[2]}_{J=2}$ with the symbol ($g$), evaluated on the $m_\pi = 391 \,\mathrm{MeV} $, $24^3\times 128$ lattice. Plotted is $e^{m_E t}\cdot C(t)$ where $m_E$ is the lowest mass state in the isoscalar $E^{++}$ sector.   \label{Eppglue}}
\end{figure} 

Following a similar construction as for the fermion bilinears in Eqn \ref{eq:bilinear}, we choose to define glueball operators using the distillation space vectors $\xi(t)$ defined in Eq.\ref{Box} on a time-slice $t$ as
\begin{equation}
\mathcal{G}(t) =\sum_{k=1}^N \xi^{k\dag}(t) \,\overleftrightarrow{D} \ldots \overleftrightarrow{D} \, \xi^k(t)
\label{eq:glueball}
\end{equation}
with up to three gauge-covariant derivatives, to give an operator of definite $J^{PC}$ at zero momentum. As in the fermion bilinears, the glueball operators are subduced into irreducible representations of the cubic group. The notable difference is that while the fermion bilinear operators defined in Eqn \ref{eq:bilinear} lead to matrices in distillation space which multiply quark propagation perambulators, the glueball operators in Eqn \ref{eq:glueball} are a trace in distillation space. Thus, the contribution of glueball operators to correlation functions is only through product of traces, such as
\begin{align}
\mathcal{D}^{\ell g}(t',t) &= \mathrm{Tr}\big[ \Phi(t') \tau^\ell(t',t') \big]\;  \mathcal{G}(t) ,\nonumber \\
\mathcal{D}^{g g}(t',t) &=  \mathcal{G}(t') \;  \mathcal{G}(t) ,\nonumber 
\end{align}
which is similar to the products of traces of quark bilinear operators occurring in the disconnected parts of Fig.~\ref{wicks}.

In Figure \ref{Eppglue} we show a sample set of correlation functions in the irrep $E^{++}$ (which should be dominated by $2^{++}$ states) using the operators $\big(\rho \times D^{[1]}_{J=1}\big)^{J=2}$, which resembles a $q\bar{q}\;^3\!P_2$ construction in the non-relativistic limit, for both light ($\ell$) and strange ($s$) quark flavors, and the glueball operator $D^{[2]}_{J=2}$ ($g$). While there are clearly non-zero signals connecting the $q\bar{q}$ and glue sectors at early times, these signals have degraded into noise by time-slice~15. These glueball correlation functions can be included in the variational analysis; however, the quality of the signals is not sufficiently precise to have a discernible impact upon the levels in the variationally determined spectrum. Rather stronger signals are observed in the $A_1^{++}$ irrep, which we do not report upon in this paper -- we require inclusion of multi-particle operator constructions before accurate determination of the scalar spectrum is feasible. For the results presented in this paper, we will not include the glueball operators in the operator basis.

\section{Extraction of the spectrum} \label{spec}

Our approach to extracting the spectrum requires computation of matrices of isoscalar correlation functions using the large basis of operators sketched in the previous section. For a given irrep of the cubic lattice symmetry we include operators subduced from $J \le 4$, with each operator appearing twice, once with the $\ell$ flavor construction and once with $s$, as defined in Eqn \ref{massbasis}.

The matrix of correlation functions is analyzed using a variational approach \cite{Michael:1985ne,Luscher:1990ck,Blossier:2009kd} which involves solving the generalized eigenvalue problem,
\begin{equation}
C(t) v^\mathfrak{n} = \lambda_\mathfrak{n}(t) C(t_0) v^\mathfrak{n}, \label{var}
\end{equation}
for eigenvalues, or ``principal correlators", related to the energy spectrum, ${\lambda_\mathfrak{n}(t) \sim e^{-E_\mathfrak{n} (t-t_0)}  }$, and eigenvectors, $v^\mathfrak{n}$, from which we can determine the spectral overlaps, $\big\langle \mathfrak{n} \big| \mathcal{O}_i^\dag(0) \big| 0 \big\rangle$. For details of our implementation of this method see Refs \cite{Dudek:2007wv, Dudek:2010wm}. 

We have previously analyzed isovector correlation functions using the same basis of subduced fermion bilinear operators~\cite{Dudek:2009qf,Dudek:2010wm}, where we found that each state in the observed spectrum had strong overlap only onto operators subduced from a single $J$ value. Furthermore, near-degenerate states were observed distributed across lattice irreps in precisely the manner expected for the distribution of $J_z$ components of a spin-$J$ state. In this study of the isoscalar spectrum we observe the same separation of states in spin, and we are able to assign a definite $J$ to each state using the methods described in~\cite{Dudek:2010wm}. 

Determination of the values of the spectral overlaps for ``hidden light", $Z^\mathfrak{n}_{\ell} = \big\langle \mathfrak{n} \big| \mathcal{O}^{\ell\dag}(0) \big| 0 \big\rangle$, and ``hidden strange", $Z^\mathfrak{n}_{s} =\big\langle \mathfrak{n} \big| \mathcal{O}^{s\dag}(0) \big| 0 \big\rangle$, allows us to investigate the flavor structure of the isoscalar meson spectrum. A particularly simple case follows if we assume two nearby states in the spectrum are orthogonal admixtures of two basis states of flavor structure $\big| \ell \big\rangle \equiv \frac{1}{\sqrt{2}} \Big( \big| u\bar{u} \big\rangle + \big| d\bar{d} \big\rangle \Big)$ and $\big|s\big\rangle \equiv \big|s\bar{s}\big\rangle$. Under this hypothesis we can write, for the two eigenstates, $\big|\mathfrak{a}\big\rangle$, $\big|\mathfrak{b}\big\rangle$, where $m_\mathfrak{a} < m_\mathfrak{b}$,
\begin{equation}
\begin{pmatrix} \big|\mathfrak{a}\big\rangle \\ \big|\mathfrak{b}\big\rangle \end{pmatrix} = \begin{pmatrix} \cos \alpha && - \sin \alpha \\ \sin \alpha && \cos \alpha \end{pmatrix}  \begin{pmatrix} \big|\ell\big\rangle \\ \big|s\big\rangle \end{pmatrix}, 
\end{equation}
and assuming that 
\begin{align}
&\big\langle \ell \big| \mathcal{O}^{\ell\dag}(0) \big| 0 \big\rangle = Z_\ell     &\big\langle s \big| \mathcal{O}^{\ell\dag}(0) \big| 0 \big\rangle &= 0 \nonumber \\
&\big\langle \ell \big| \mathcal{O}^{s\dag}(0) \big| 0 \big\rangle = 0       &\big\langle s \big| \mathcal{O}^{s\dag}(0) \big| 0 \big\rangle &= Z_s, \nonumber
\end{align}
we can determine the mixing angle, $\alpha$, from the combination 
\begin{equation}
	\tan \alpha = \sqrt{ - \frac{Z^{\mathfrak{b}}_{\ell} \, Z^{\mathfrak{a}}_{s} }{Z^{\mathfrak{a}}_{\ell} \, Z^{\mathfrak{b}}_{s}} } .	 \label{angle}
\end{equation}
We can actually get one determination of $\alpha$ for each operator construction, $\mathbf{\Gamma}$, and we expect the determinations to agree since the dependence on $Z_\ell, Z_s$, which will vary operator-to-operator, has canceled in the particular ratio formed. 

\subsection{Variational analysis of $T_1^{++}$ }
\begin{figure*}
 \centering
\includegraphics[width=.83\textwidth
]{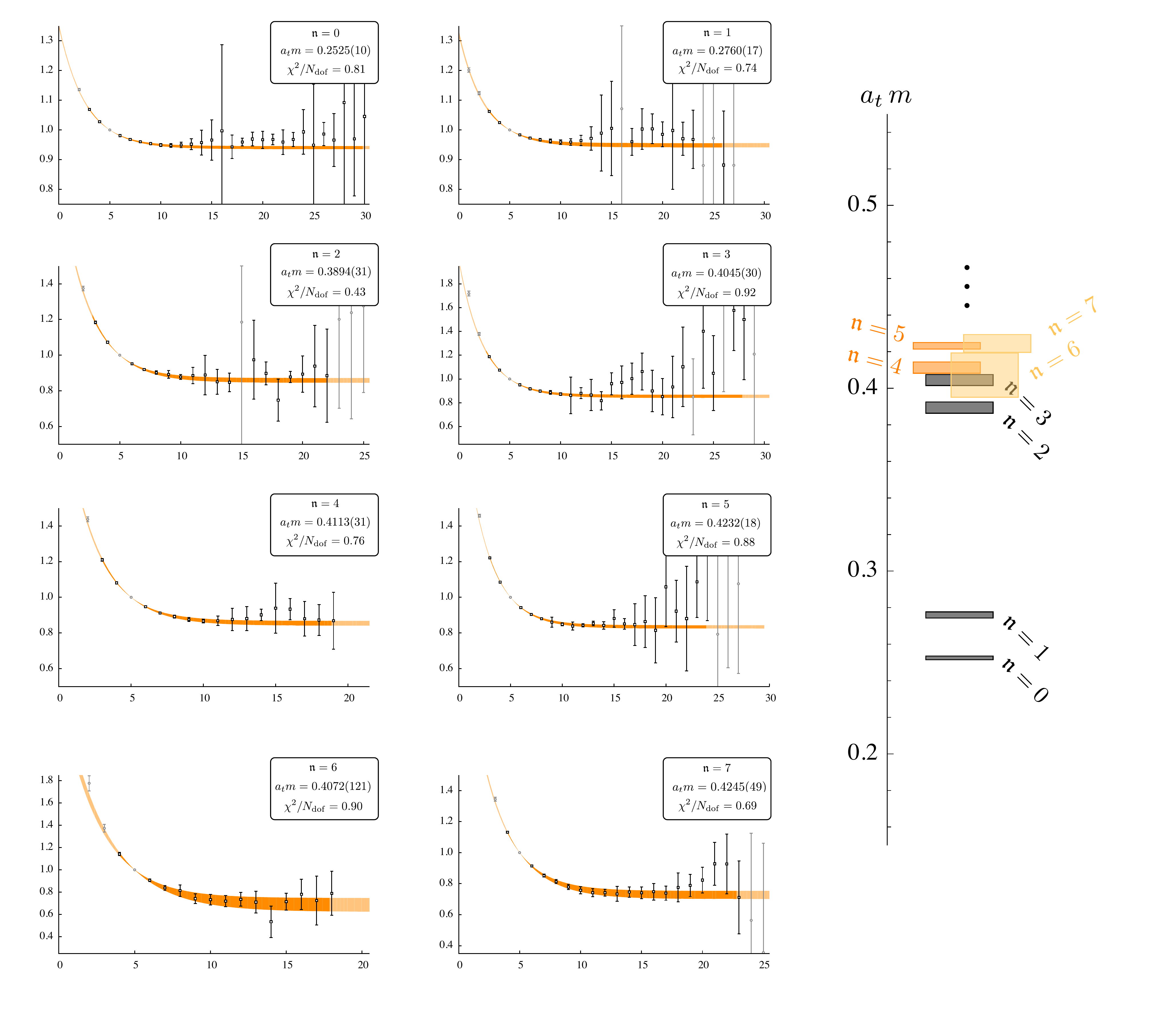} 
\caption{Lowest eight ``principal correlators" of the $T_1^{++}$ correlation matrix at $m_\pi = 391\,\mathrm{MeV}$ on the $24^3\times 128$ lattice. Plotted is $\lambda_\mathfrak{n}(t) \cdot e^{m_\mathfrak{n} (t-t_0)}$ with $t_0/a_t=5$,  and the fits to the data using the form $\lambda_\mathfrak{n}(t) = (1-A_\mathfrak{n})\, e^{-m_\mathfrak{n} (t-t_0)}     +A_\mathfrak{n} e^{-m'_\mathfrak{n} (t-t_0)} $ with $A_\mathfrak{n} \ll 1$ and $m'_\mathfrak{n}  \gg m_\mathfrak{n}$. Light grey points are not used in the fit.
 \label{T1pp_princorrs}}
\end{figure*}

As an example of the procedure outlined above, we present here the result of variational analysis of the $T_1^{++}$ correlator matrix on the $m_\pi = 391\,\mathrm{MeV}$ lattice of volume $24^3\times 128$. In this case we used 18 operators subduced from $J=1$ (the same 9 operators in each of $\ell$ and $s$ flavor constructions), two $J=3$ operators and two $J=4$ operators, leading to a matrix of size $22 \times 22$. Figure \ref{T1pp_princorrs} shows the resulting lowest eight principal correlators for the case $t_0/a_t = 5$ and the fits to the time-dependence which yield estimates of the mass spectrum. Note that the spectrum shows near-degenerate states which are being cleanly extracted -- this is possible because of the orthogonality inherent in the variational method.

\begin{figure*}
 \centering
\includegraphics[width=.93\textwidth
]{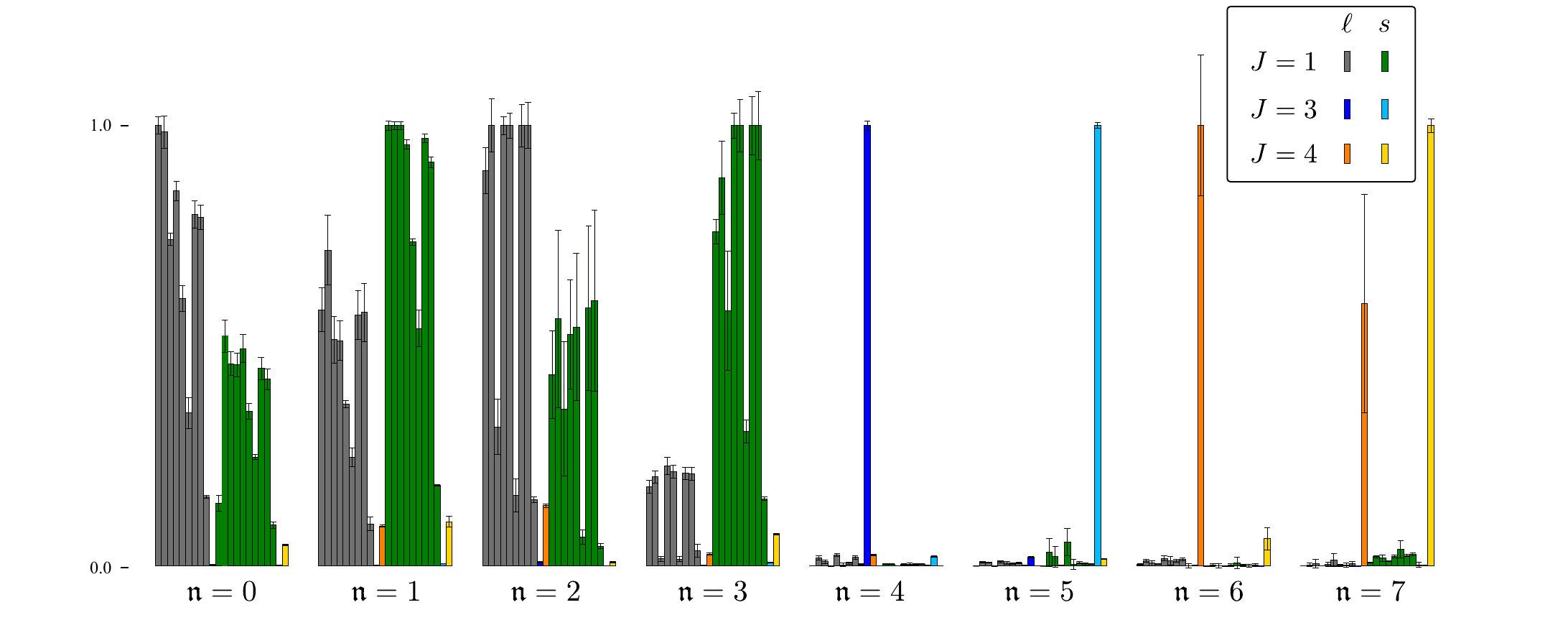} 
\caption{Overlap matrix elements $Z^\mathfrak{n}_{(\ell,s)i}  = \big\langle \mathfrak{n} \big| \mathcal{O}^{(\ell,s)\dag}_i \big| 0 \big\rangle$ for the lightest eight states of the $T_1^{++}$ correlation matrix at $m_\pi = 391\,\mathrm{MeV}$ on the $24^3\times 128$ lattice. Each bar represents a distinct operator construction, $\mathbf{\Gamma}$, with the color-coding indicating the $J$ from which the operator was subduced, and the flavor construction of the quark fields. The bars are normalized such that for each operator the largest overlap onto any state in the extracted 22 state spectrum is $1.0$.
\label{T1pp_histo}}
\end{figure*}

Figure \ref{T1pp_histo} shows the relative size of overlaps $\big\langle \mathfrak{n} \big| \mathcal{O}^{(\ell,s)\dag}_i \big| 0 \big\rangle$ for these lowest eight states. Clearly we can identify the lowest four states as being $J=1$, the next two as $J=3$ and the next two as $J=4$. It is also clear that the $J=1$ eigenstates are not diagonal in the flavor space, with each being an admixture of $\big|\ell \big\rangle$ and $\big| s \big\rangle$, while the $J=3$ states do appear to be diagonal in this flavor space.

Motivated by the relative similarity in mass of the states $\big|\mathfrak{n}=0 \big\rangle$ and $\big|\mathfrak{n}=1 \big\rangle$ and the similarly between the histograms in Figure \ref{T1pp_histo} with $\ell \leftrightarrow s$ as $(\mathfrak{n}=0) \leftrightarrow (\mathfrak{n}=1)$, we will proceed with the hypothesis that these two states are simple admixtures of  $\big| \ell \big\rangle $ and $\big|s\big\rangle$ basis states as described earlier. 

\begin{figure}
 \centering
\includegraphics[width=.49\textwidth
]{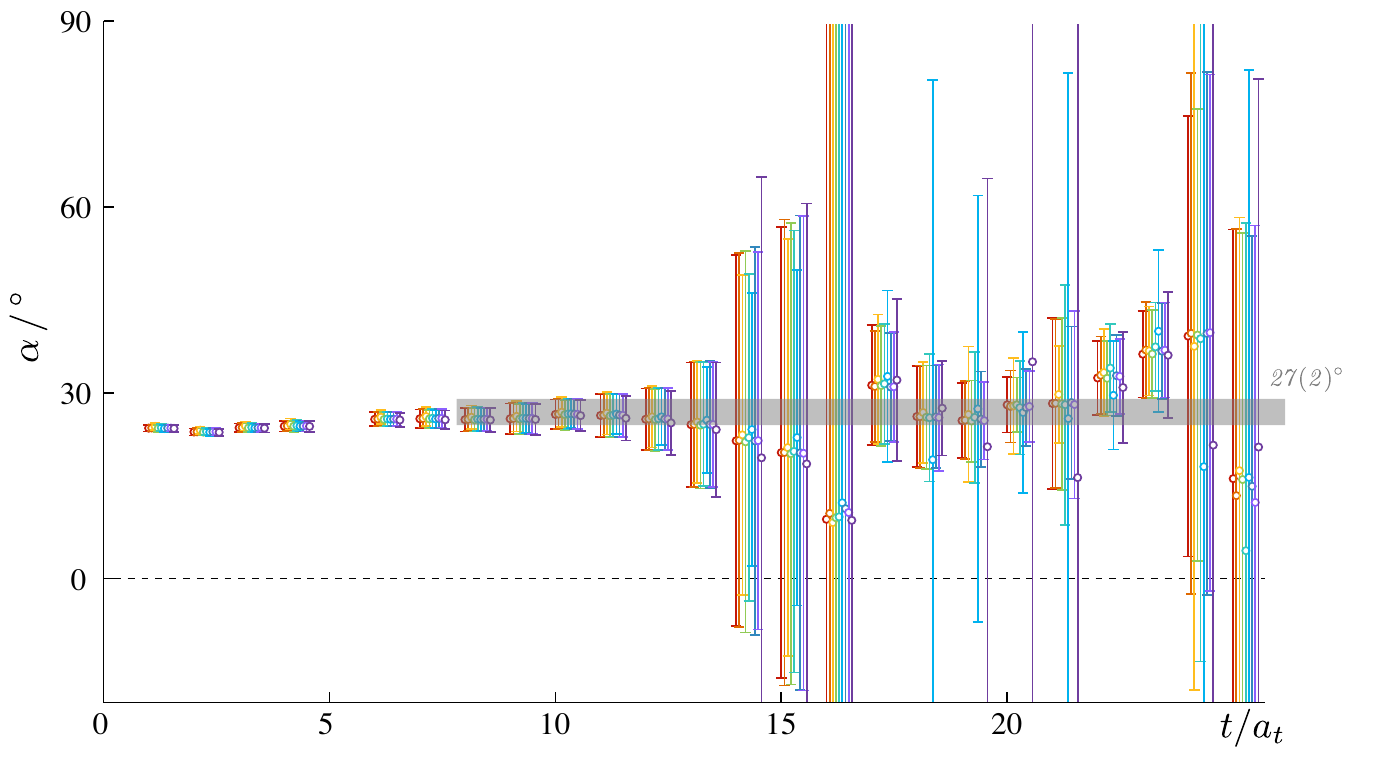} 
\caption{$\alpha(t)$ from Eqn \ref{angle} for 9 different constructions, $\mathbf{\Gamma}$ (different colors), for the lowest two states $|\mathfrak{n}=0\rangle,\, |\mathfrak{n}=1\rangle$ in $T_1^{++}$. Gray band shows a constant fit in time.  \label{f1_angles}}
\end{figure}

In our practical solution of Eqn \ref{var}, we solve independently on each timeslice, meaning that we determine $v_i^\mathfrak{n}(t)$ -- these should be constants in $t$ for $t>t_0$ if we are correctly describing the spectrum of states contributing to $C_{ij}(t)$. A simple linear transformation applied to $v_i^\mathfrak{n}(t)$ yields the overlap factors $Z^\mathfrak{n}_i(t)$ which should also be constant. The histograms in Figure~\ref{T1pp_histo} correspond to the results of constant fits to $Z^\mathfrak{n}_i(t)$. We can obtain the angle $\alpha$ in Eqn~\ref{angle} as a function of $t$ using the $Z^\mathfrak{n}_i(t)$ and it should also be constant, as it is seen to be in Figure~\ref{f1_angles} for the lowest two states $\big| \mathfrak{n}=0\big\rangle,\, \big| \mathfrak{n}=1\big\rangle$. As we can see, the determined angle is consistent for all the $J=1$ subduced operators used in the basis, and a constant fit to all the data yields our best estimate of $\alpha = 27(2)^\circ$.

\subsection{Interpretation of the spectrum} \label{interpret}

The extracted isoscalar spectrum across irreps shows the same sort of systematics in mass degeneracy and spectral overlaps that we observed in the isovector spectrum extracted using the same fermion bilinear operator constructions~\cite{Dudek:2010wm}. In particular the distribution of states across irreps is compatible with subduction of \mbox{``single-hadron"} states of definite $J$. This separation of states according to continuum spin suggests that mesons which fit within the lattice volume are not resolving the cubic symmetry of the fine lattice grid~\cite{Davoudi:2012ya}. 

But this simple pattern of states is \emph{too} simple, and indicates that in these calculations we are not resolving the complete spectrum of eigenstates of QCD in a finite cubic volume.

Within a finite cubic volume, there should be a discrete spectrum of multi-hadron states, which, if hadrons did not interact, would lie at predictable energies. For example, there would be a spectrum of two-hadron states, $a_t E_{AB} = \sqrt{(a_t m_A)^2 + n_A^2 \left(\tfrac{2\pi}{\xi L_s/a_s}\right)^2 } +  \sqrt{(a_t m_B)^2 + n_B^2 \left(\tfrac{2\pi}{\xi L_s/a_s}\right)^2 }$, where $n_A^2, n_B^2$ are integers, related to the momenta of the hadrons, $A,\,B$, with allowed values which depend upon the irrep under consideration. This spectrum of states has a strong volume dependence and a distribution across irreps that is quite different to the subduction of a single hadron of spin $J$ \cite{Dudek:2012gj}. 

In fact, the true eigenstates will be volume-dependent admixtures of basis states with good overlap onto ``single hadron" operators and multi-hadron states. An explicit example can be seen in Ref.~\cite{Dudek:2012xn}. Figure 1 therein presents the isovector $T_1^{--}$ spectrum in finite-volume extracted from variational analysis using firstly just fermion bilinear operators, and secondly using in addition a set of operators which resemble a pair of pions with relative momentum. In the latter case there are clearly extra states present which strongly resemble pairs of pions, but also the lowest lying state that we identify with the $\rho$ (when computing with only single-hadron operators) becomes two states which lie within an energy region corresponding to the hadronic width of the $\rho$ resonance. Analysis of these levels using the L\"uscher technique \cite{Luscher:1990ux} allows us to resolve the $\pi\pi$ scattering amplitude and identify a rapidly rising phase-shift that we associate with the $\rho$ resonance. We argue that the presence of a single low-lying level when only fermion-bilinears are used suggests the existence of a single low-lying narrow meson, as confirmed in the more complete analysis. 

Based upon the more careful analysis of the $\rho$ described above we suggest that the spectra presented herein for the isoscalar meson spectrum are likely a reasonable guide to the resonant state content of the infinite-volume theory at these pion masses. There may be broad structures that we are missing, and the ``masses" extracted for unstable states should not be considered precision estimates, but can probably be assumed to lie somewhere within the hadronic width of the meson. Ultimately extraction of the full resonant content of the excited meson spectrum will require more sophisticated calculations including relevant multi-meson-like operator constructions. 

In the results presented in the next section, we will exclude any consideration of the $0^{++}$ sector which has a notoriously complicated phenomenology -- experimentally it features a large number of low-lying resonances on top of a significant broad structure due to the $\sigma$ resonance. We suspect that understanding this channel will demand the kind of sophisticated calculation suggested above in which meson-meson operators for a set of channels ($\pi\pi$, $K\overline{K}$, $\eta\eta$ \ldots) are included.

Before we embark upon a presentation of the large set of results obtained, we remind the reader that in these calculations isospin is an exact symmetry and electromagnetic effects are not present.

\section{Results} \label{results}

Here we present the spectra obtained, using the methods described previously, on the lattices listed in Table \ref{tab:lattices}.

\subsection{$\mathbf{m_{\boldsymbol{\pi}} \boldsymbol{=} 391\, MeV}$, $\mathbf{24^3 \boldsymbol{\times} 128}$}
\begin{figure*}
 \centering
\includegraphics[width=.99\textwidth
]{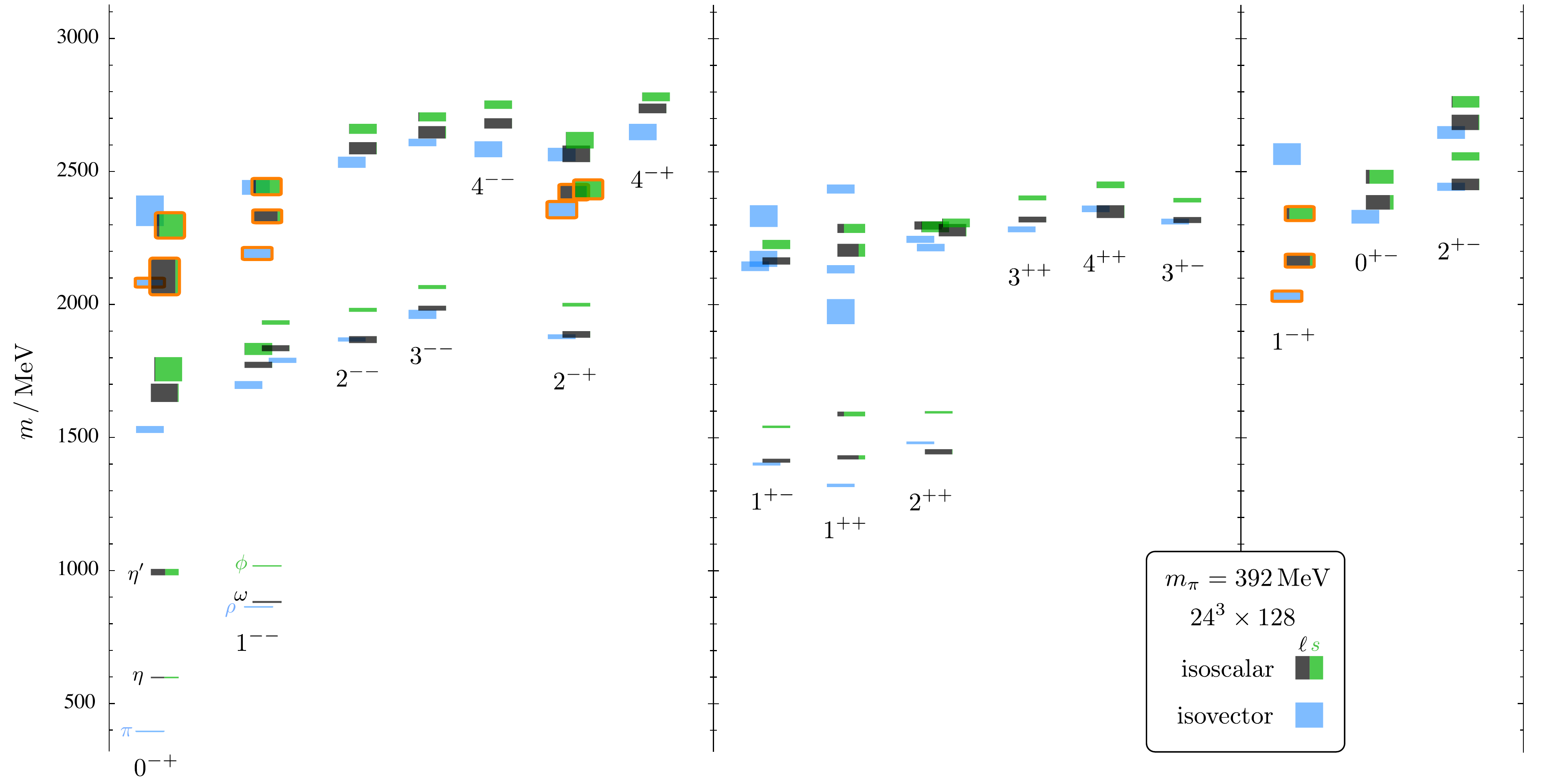} 
\caption{Isoscalar (green/black) and isovector (blue) meson spectrum on the $m_\pi = 391\,\mathrm{MeV}$, $24^3\times 128$ lattice. The vertical height of each box indicates the statistical uncertainty on the mass determination. States outlined in orange are the lowest-lying states having dominant overlap with operators featuring a chromomagnetic construction -- their interpretation as the lightest hybrid meson supermultiplet will be discussed later.\label{840_V24}}
\end{figure*}

We can illustrate the general trend of spectrum results using this case, the lowest quark mass and largest volume considered. In Figure \ref{840_V24} we show the isoscalar and isovector spectrum separated by $J^{PC}$. The black/green boxes indicate the degree of hidden light -- hidden strange mixing as determined by the mixing angle\footnote{for the lighter state in the mixed pair the size of the black part of the bar is $\cos^2 \alpha$ and the green $\sin^2 \alpha$ and vice-versa for the heavier state}, extracted as described earlier. We observe that with a few exceptions, notably  $0^{-+}$ and $1^{++}$, the spectrum shows very little mixing of light and strange. A detailed spectrum of exotic $J^{PC}$ mesons is extracted, as is a set of states in the non-exotic spectrum that have large overlap onto operators featuring a chromomagnetic construction -- the corresponding hybrid meson phenomenology will be discussed later.

\subsection{$\mathbf{ m_{\boldsymbol{\pi}} \boldsymbol{=} 391\, MeV} $, volume dependence}
\begin{sidewaysfigure*}
\centering
 \vspace{7cm}
\includegraphics[height=4in]{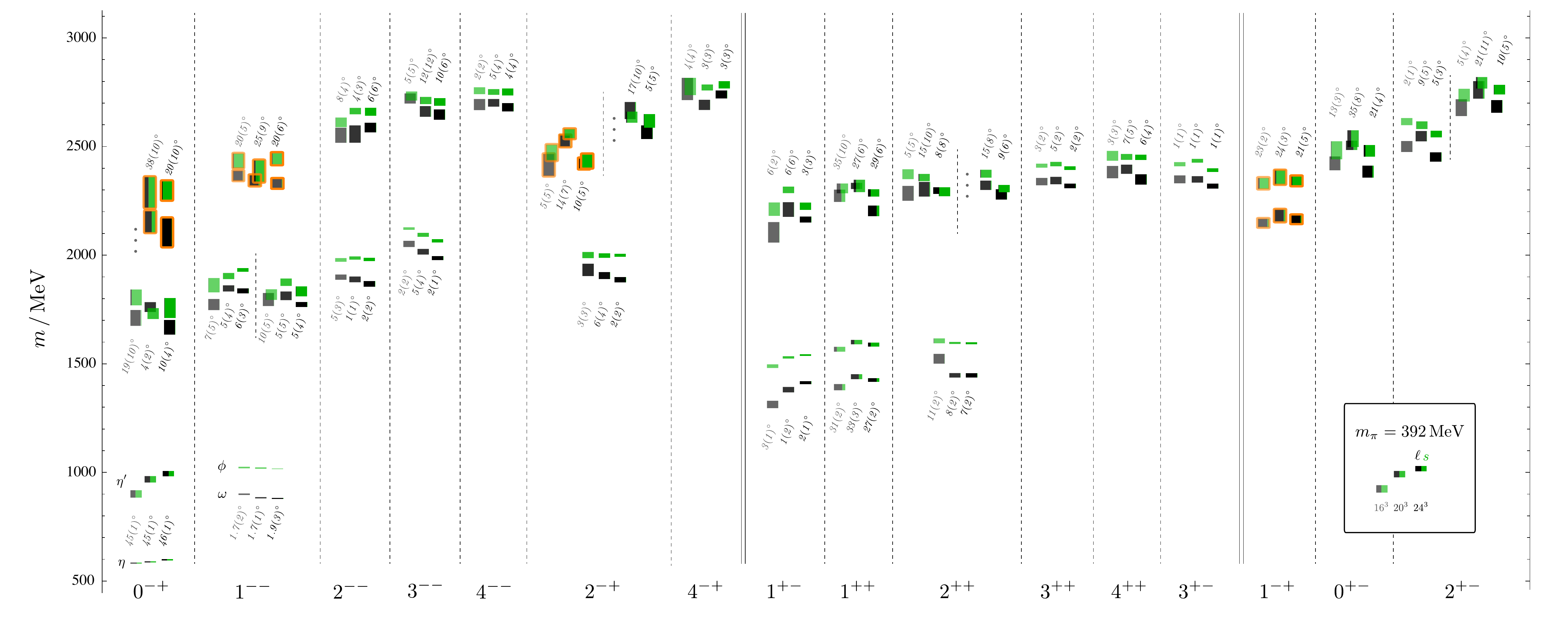} 
\caption{Volume dependence of isoscalar spectrum for $m_\pi = 391\,\mathrm{MeV}$.   \label{840_voldep}}
\end{sidewaysfigure*}

The dependence of the isoscalar spectrum on the lattice volume is presented in Figure \ref{840_voldep}.

The gross structure of the spectrum is observed to be robust with respect to changes in the volume and what changes there are tend to be not much larger than the statistical uncertainty. This relatively mild dependence on the volume should come as a surprise given that the true finite-volume spectrum will be made up of admixtures featuring strongly volume-dependent multi-meson states. We remind the reader of the discussion in Section \ref{interpret}, where we argue that the fermion bilinear operators in use here do not have good overlap onto multi-meson basis states. With this in mind, variation of an extracted energy within the hadronic width of a state as the volume is varied should not come as a surprise.

Perhaps the only unexpected variation with volume is that of the $\eta'$, which should be exactly stable in our calculations since the decay channel $\eta \pi \pi$ is kinematically closed. We do not have an explanation for this variation, although a possible candidate might be the volume dependence of gauge-field configurations of non-trivial topology since they would contribute to the $U(1)_A$ anomaly that can split the $\eta'$ from the $\eta$.

\subsection{$\mathbf{SU(3)_F}$ point, $\mathbf{m_{\boldsymbol{\pi}} \boldsymbol{=} 702\,MeV}$,  $\mathbf{(16,20)^3 \boldsymbol{\times} 128}$}

\begin{sidewaysfigure*}
\centering
 \vspace{7cm}
\includegraphics[height=5in]{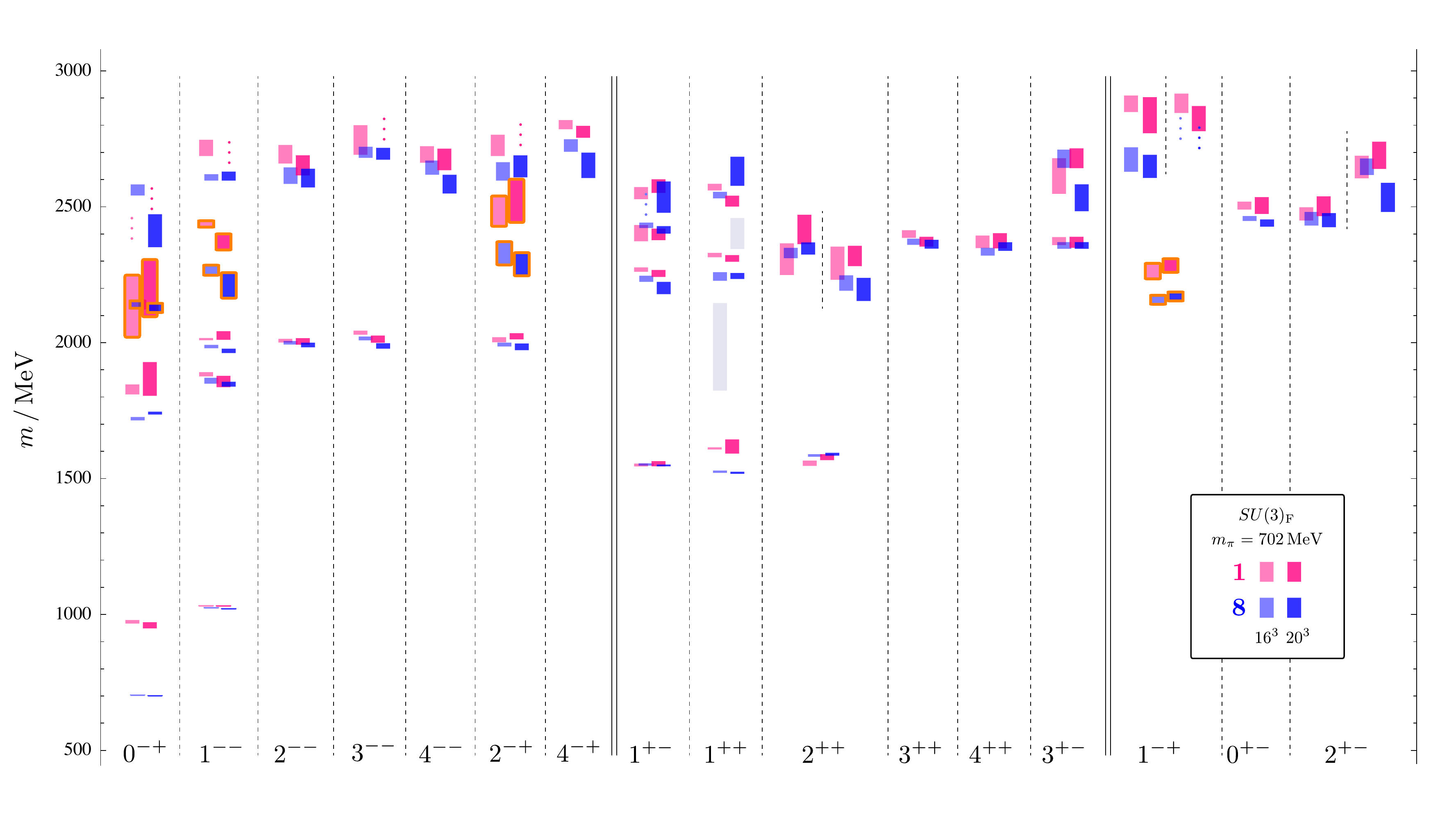} 
\caption{Octet and singlet meson spectrum on the exact $SU(3)_F$ symmetric $m_\pi = 702\,\mathrm{MeV}$, $16^3\times 128$ and $20^3 \times 128$ lattices. States outlined in orange are the lowest-lying states having dominant overlap with operators featuring a chromomagnetic construction. \label{743}}
\end{sidewaysfigure*}

In this case we take all three quark flavors to be mass degenerate, with the mass we have tuned to correspond to the physical strange quark. Here, because there is an exact $SU(3)$ flavor symmetry, we characterize mesons in terms of their $SU(3)_F$ representation, octet ($\mathbf{8}$) or singlet ($\mathbf{1}$), and compute correlation matrices using the basis in Eqn \ref{SU3basis}. The octet correlators feature only connected diagrams while the singlets receive an additional contribution from a disconnected diagram. Since the strange quarks are now no heavier than the ``light" quarks, any splitting between states in the octet and singlet spectra is purely due to the disconnected diagrams and thus to ``annihilation dynamics". In Figure \ref{743} we present the spectra extracted on two lattice volumes.

\subsection{Quark mass and volume dependence}

\begin{figure*}
 \centering
\includegraphics[width=.45\textwidth]{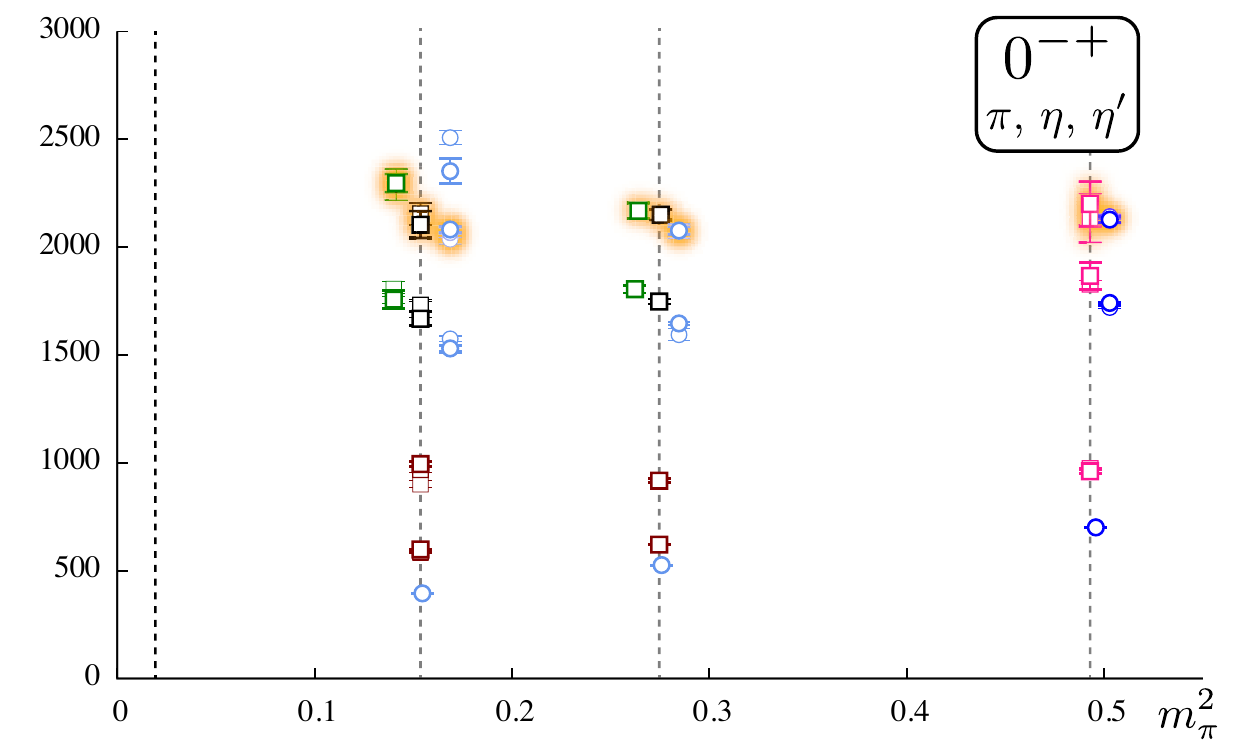} 
\includegraphics[width=.45\textwidth]{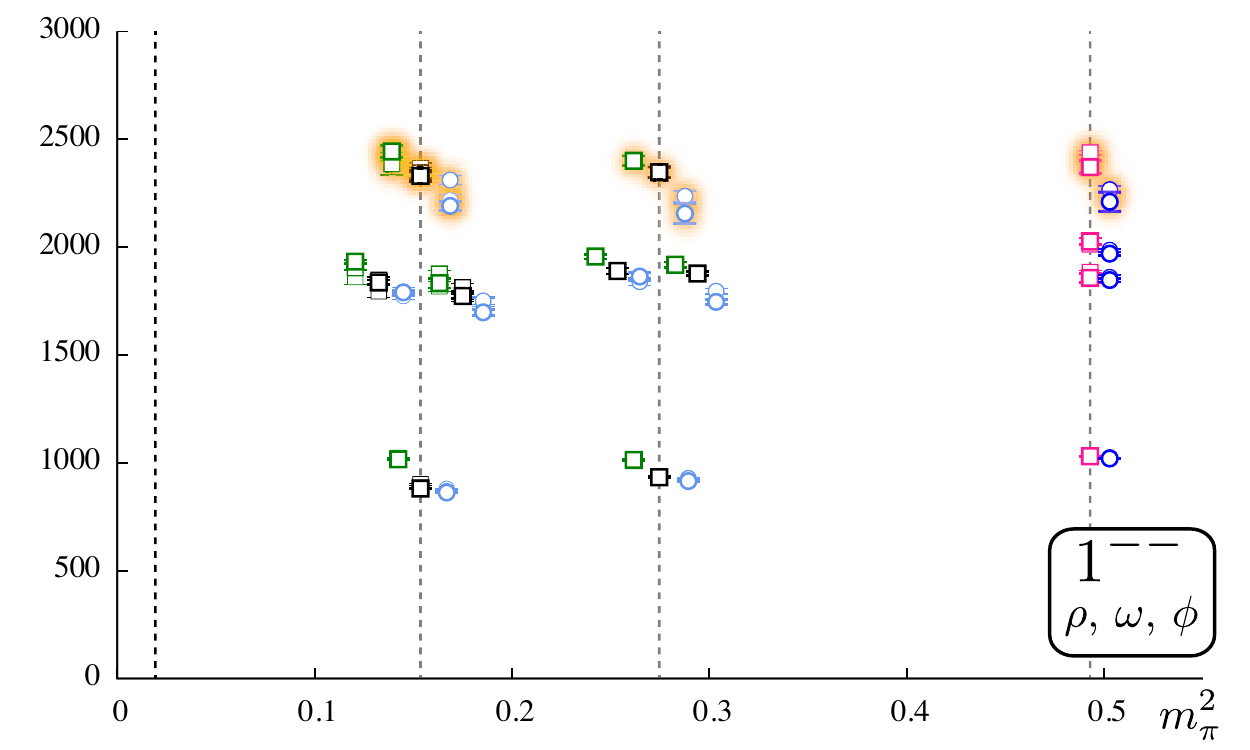} 
\caption{
Meson spectrum, in MeV, as a function of $m_\pi^2$ in $\mathrm{GeV}^2$. Isoscalar states dominantly of light quark construction shown in black, and dominantly of strange quark construction in green. States with a large degree of light/strange mixing shown in dark red. Isovector states shown in light blue. Many points are displaced horizontally from the correct value of $m_\pi^2$ (shown by the vertical dashed lines) for clarity. With exact $SU(3)_F$ at the largest quark mass, flavor octet states shown in blue and flavor singlet states in pink. At each pion mass, the thickest lined points shows the largest volume available, with thinner lines representing smaller volumes -- note that in the $m_\pi = 524\,\mathrm{MeV}$ case there are two volumes ($16^3, 20^3$) for the isovector spectrum, but only one ($16^3$) for the isoscalar. States surrounded by an orange glow are those found to have large overlap onto chromomagnetic operators and which are suspected to lie in the lightest hybrid meson supermultiplet.
\label{0mp1mm}}
\end{figure*}

\begin{figure*}
 \centering
\includegraphics[width=.43\textwidth]{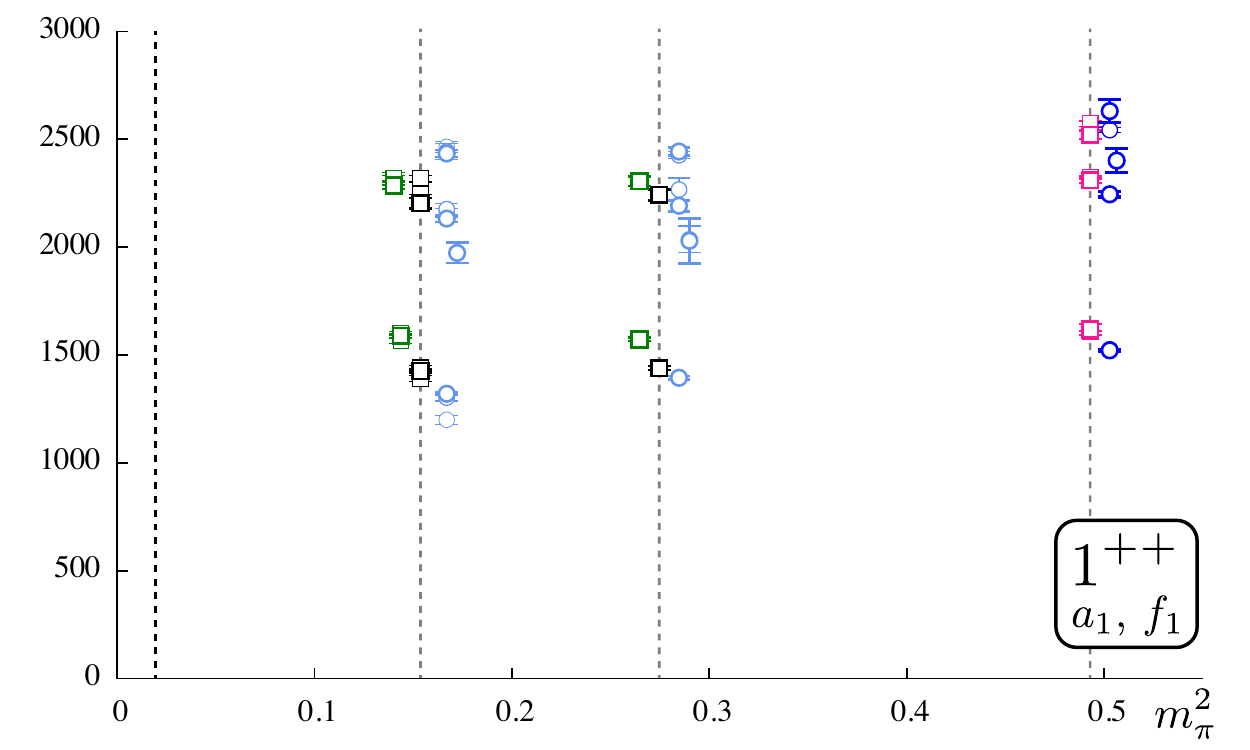} 
\includegraphics[width=.43\textwidth]{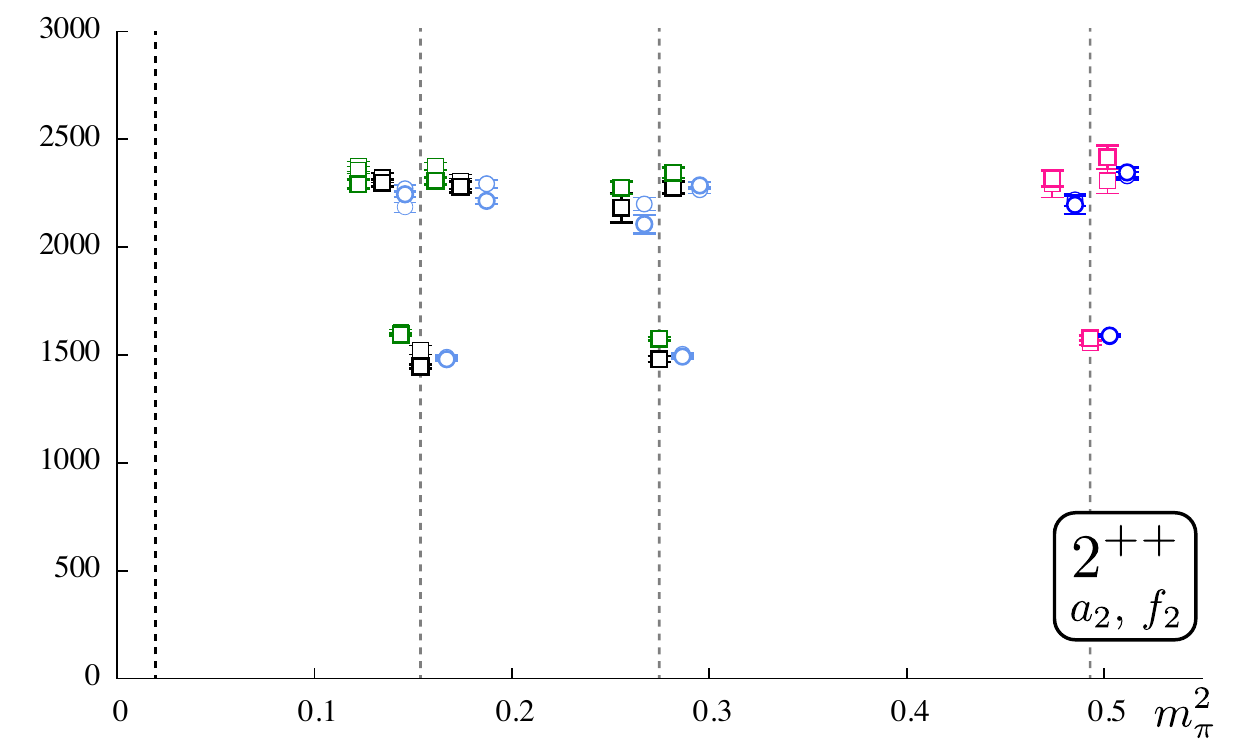}
\includegraphics[width=.43\textwidth]{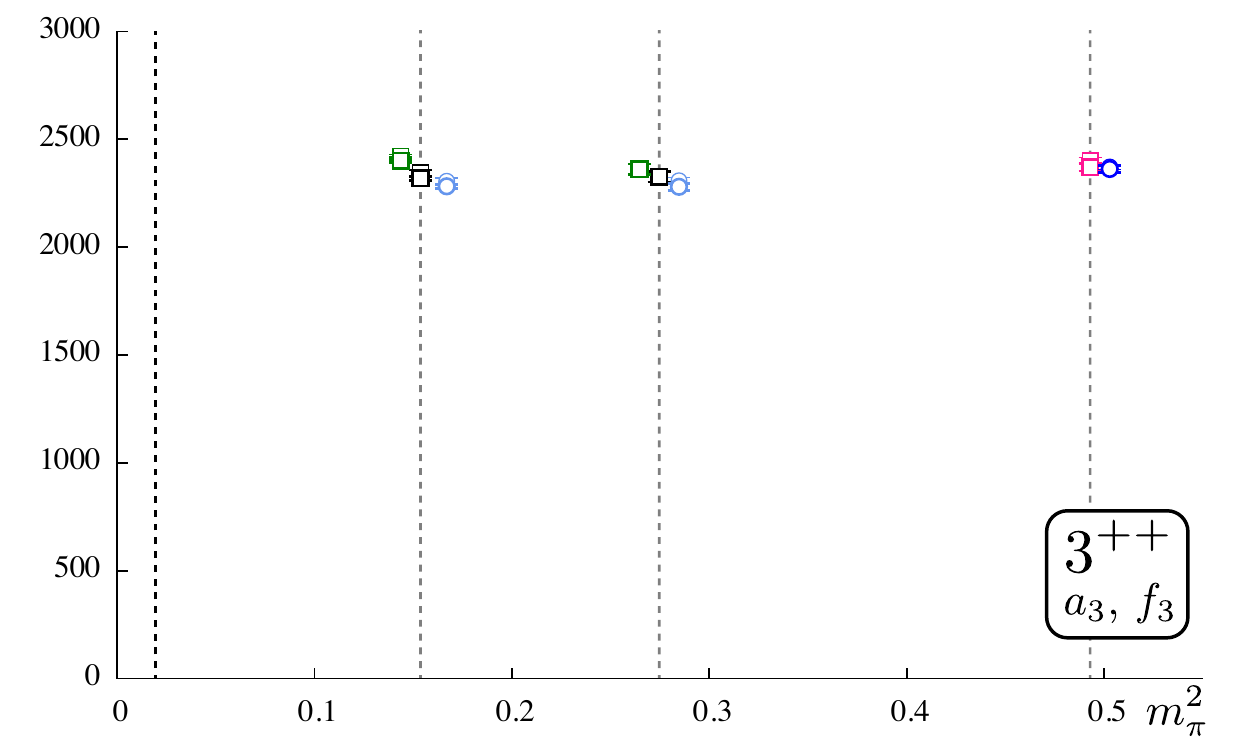}
\includegraphics[width=.43\textwidth]{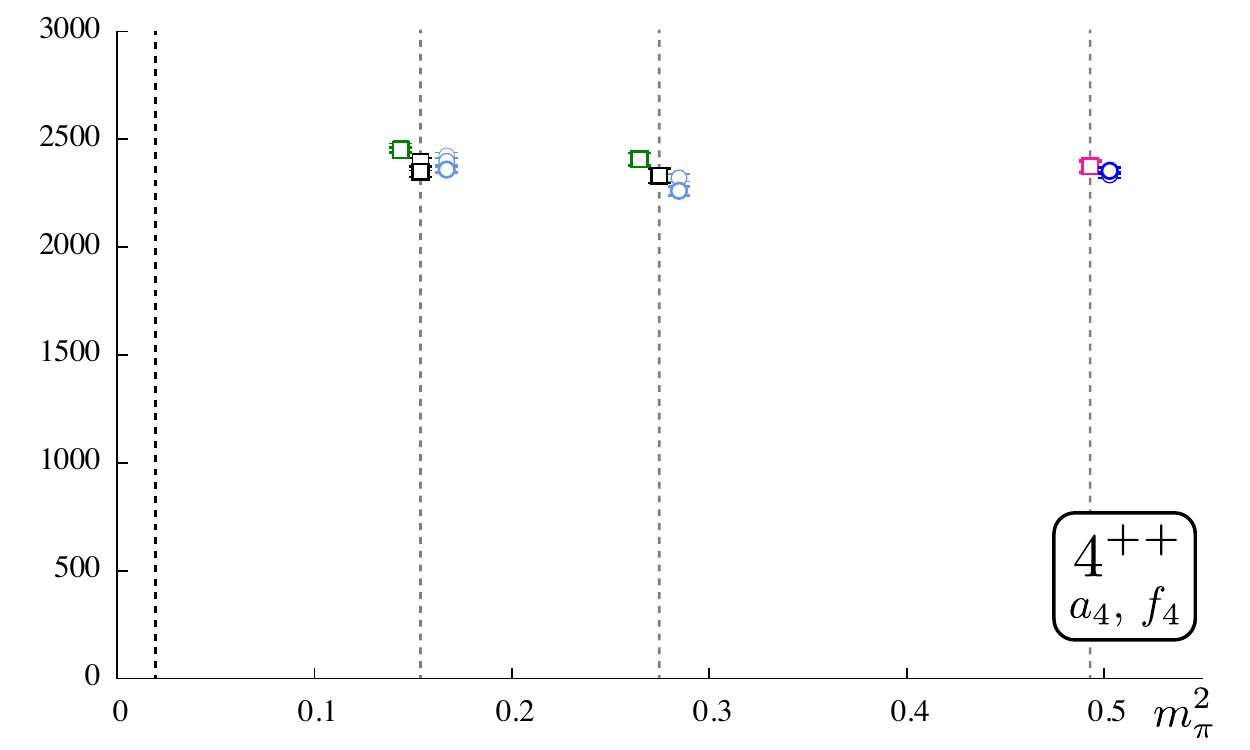}   
\caption{See caption of Figure \ref{0mp1mm}\label{Jpp}}
\end{figure*}

\begin{figure*}
 \centering
 \includegraphics[width=.45\textwidth]{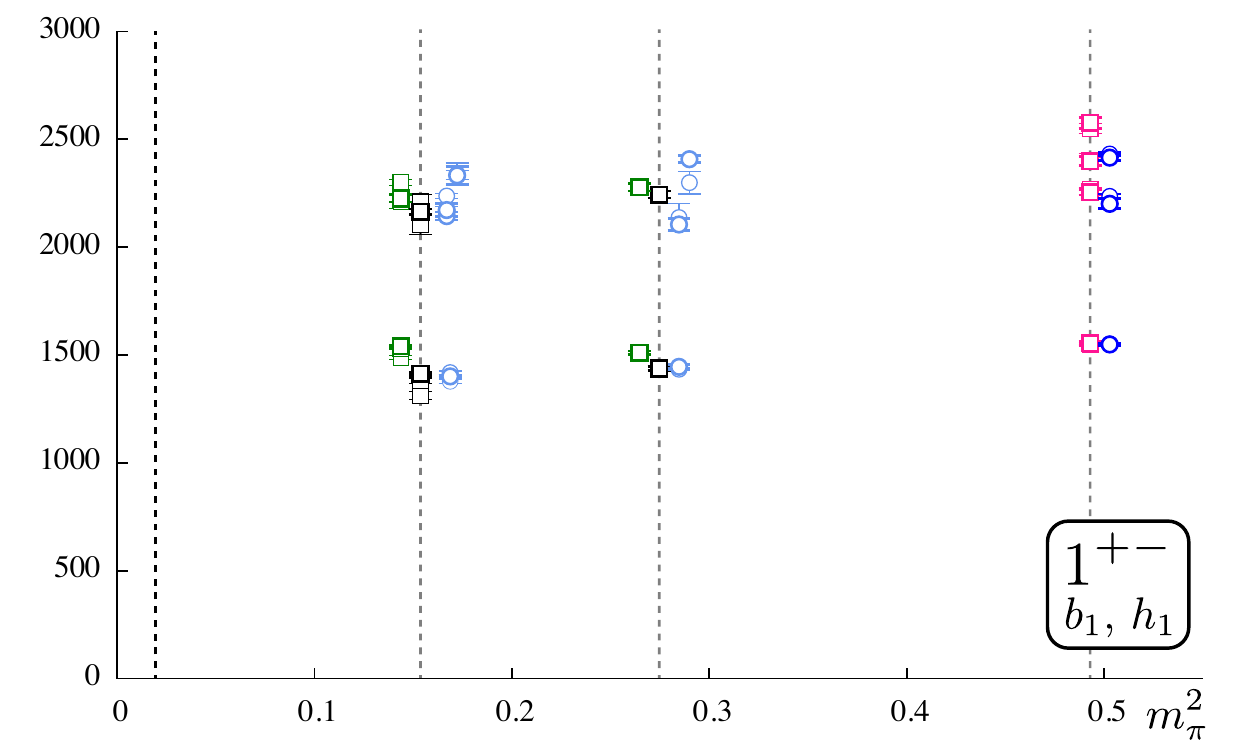} 
\includegraphics[width=.45\textwidth]{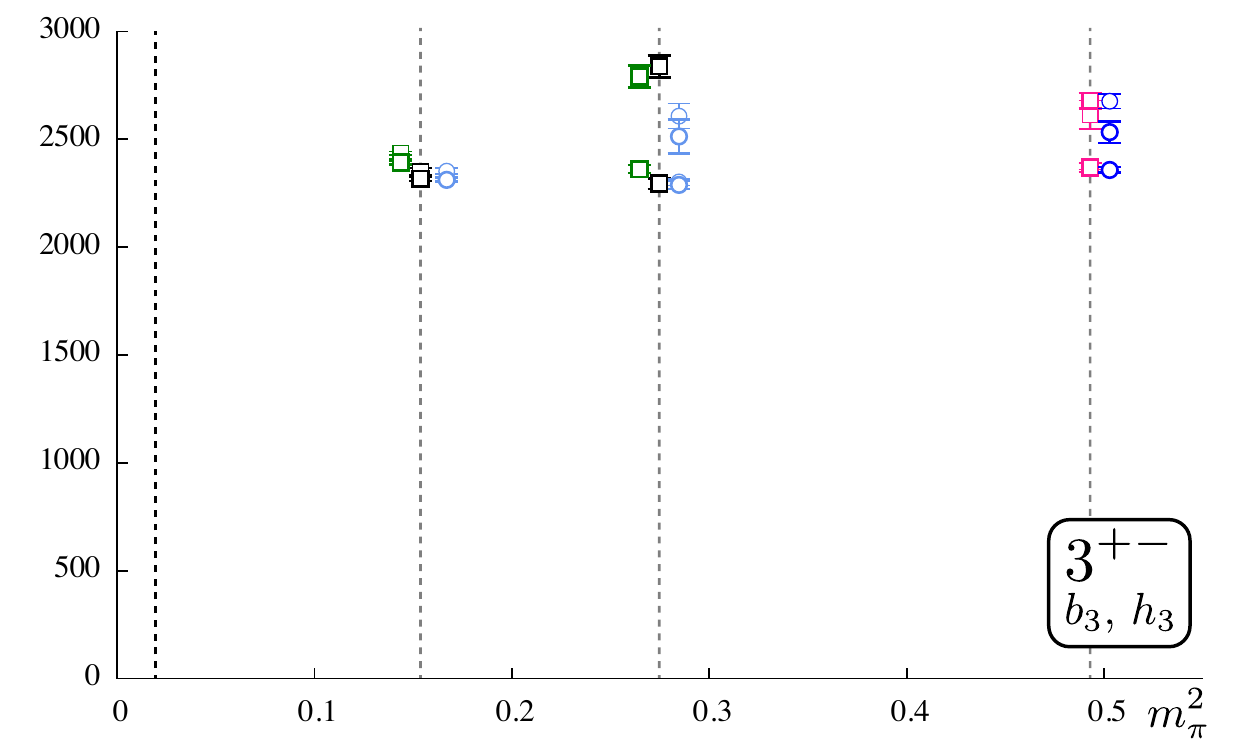} 

\vspace{15mm} 
 
\includegraphics[width=.45\textwidth]{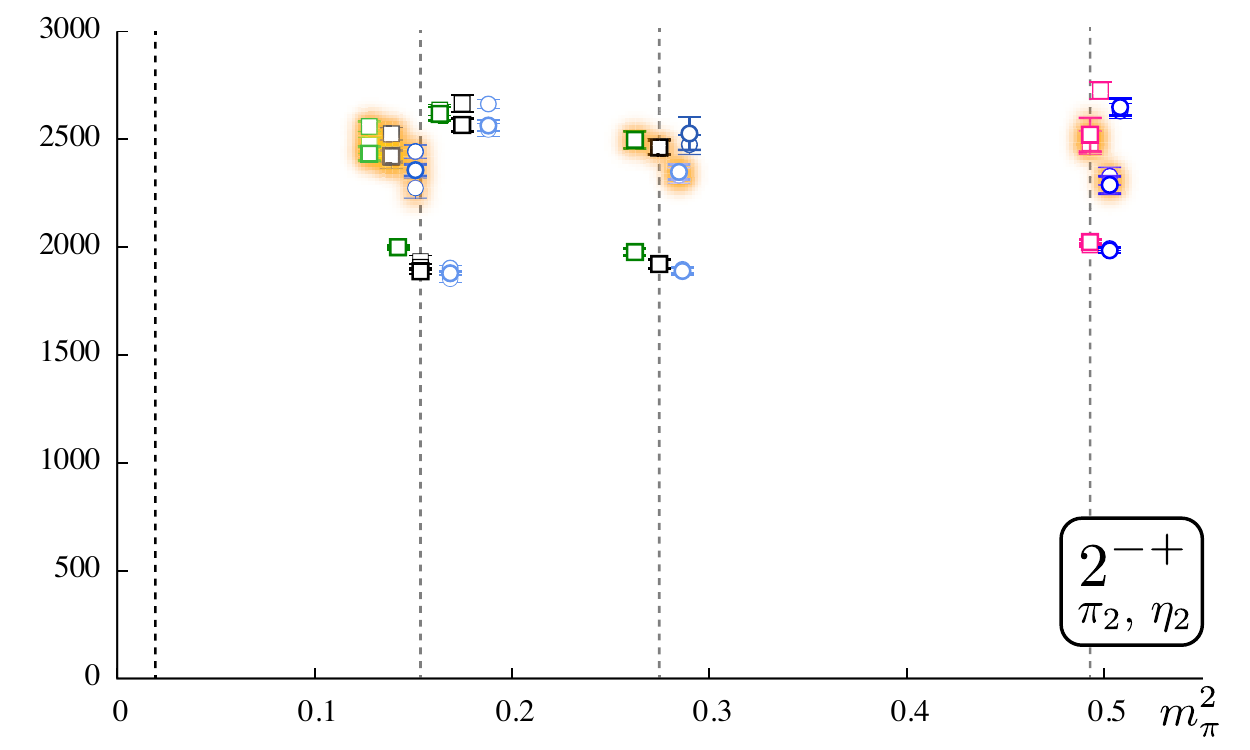}
\includegraphics[width=.45\textwidth]{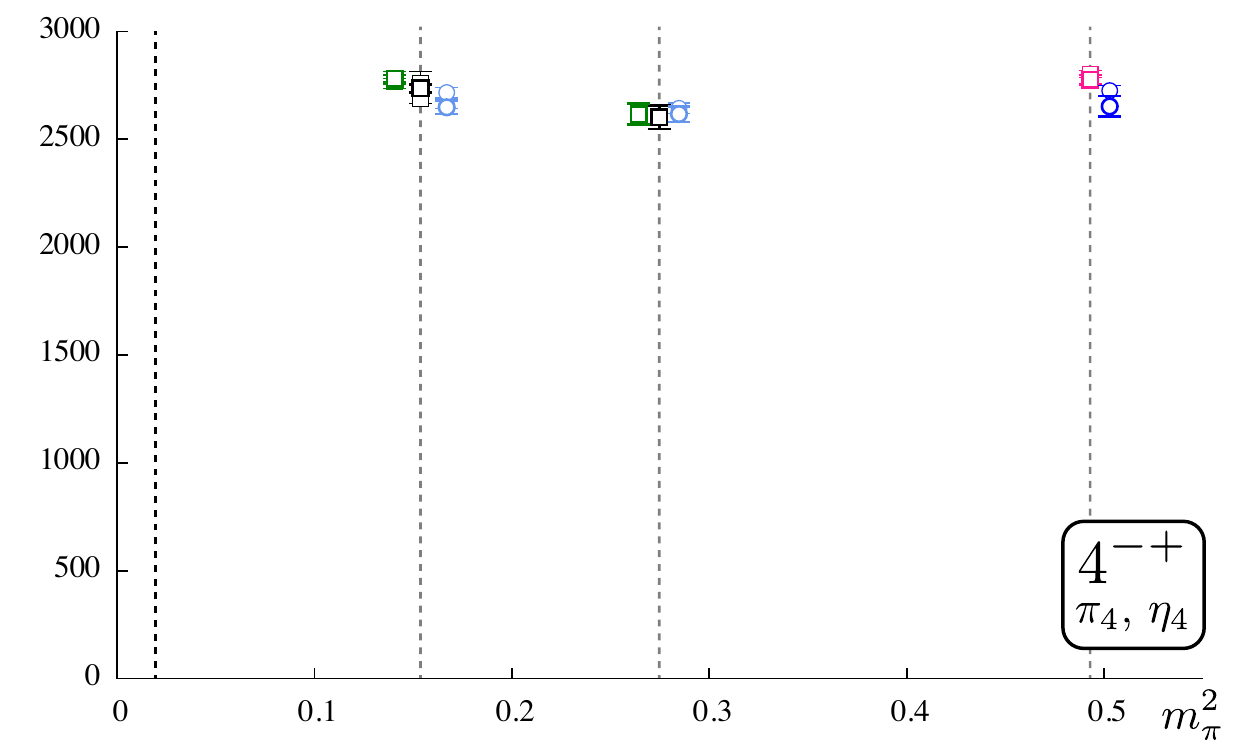}  

\vspace{15mm}

\includegraphics[width=.32\textwidth]{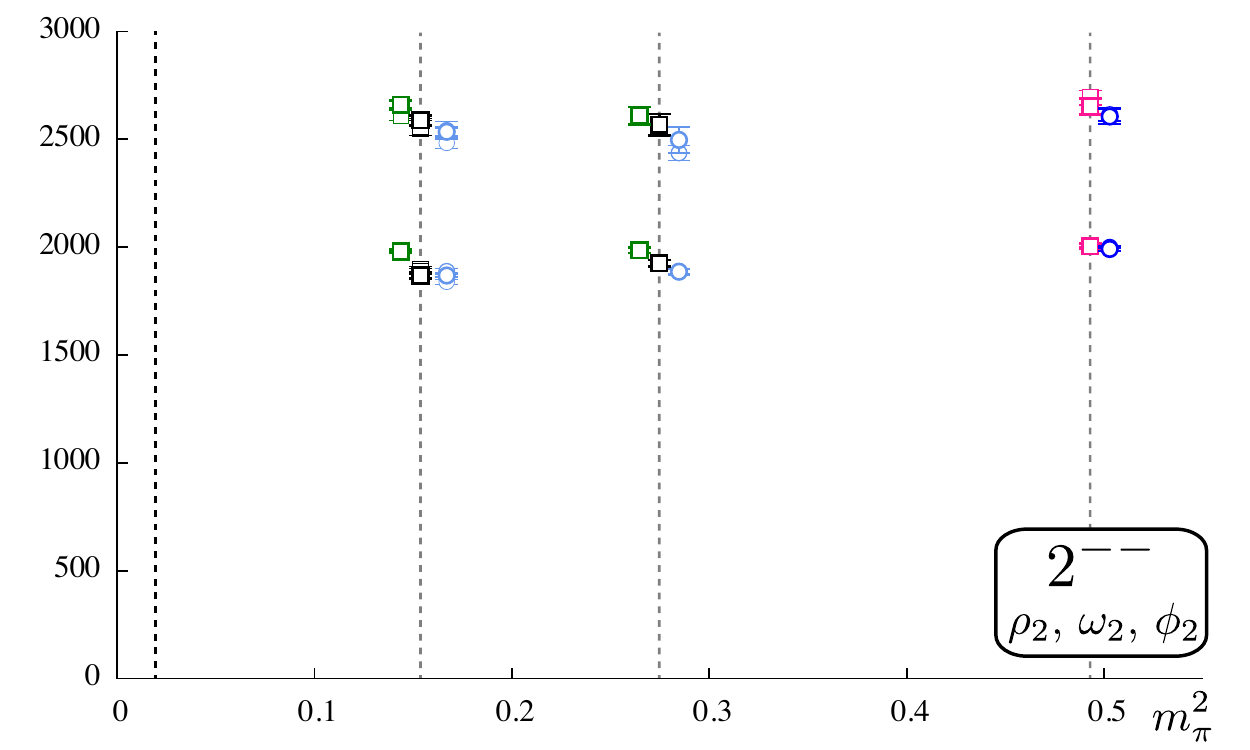} 
\includegraphics[width=.32\textwidth]{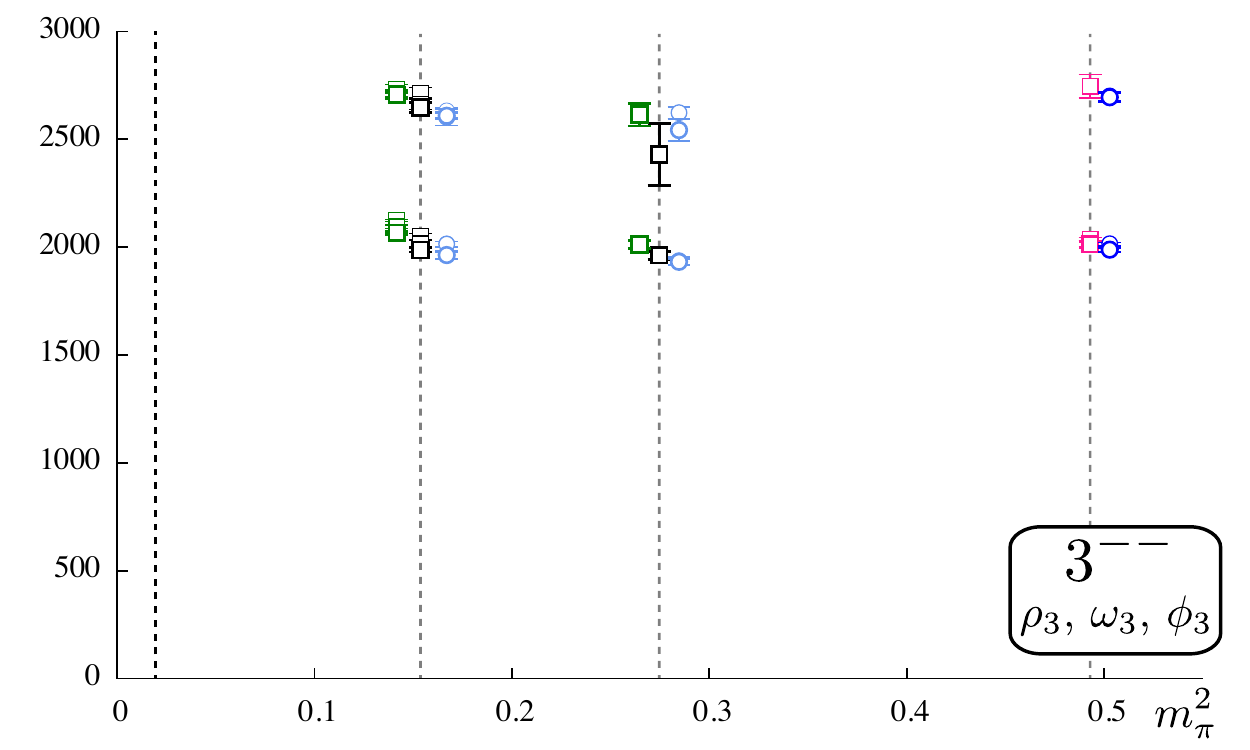} 
\includegraphics[width=.32\textwidth]{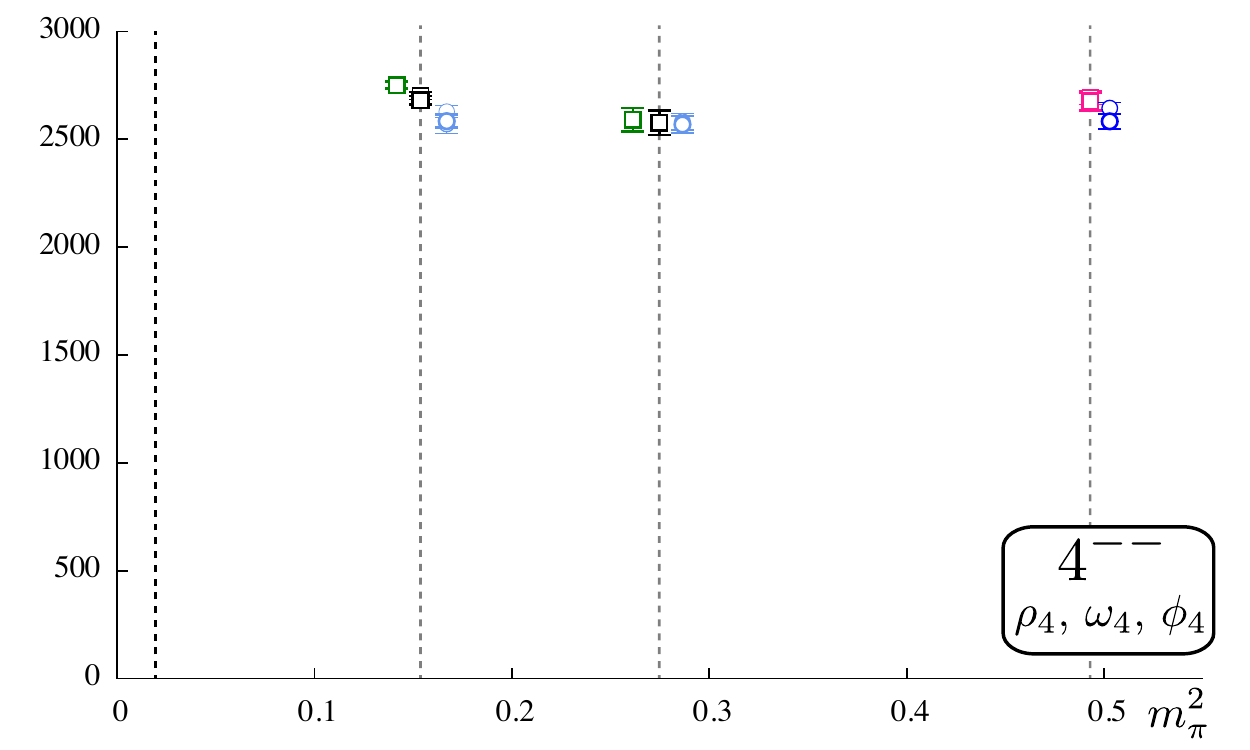} 

\vspace{15mm}

\includegraphics[width=.32\textwidth]{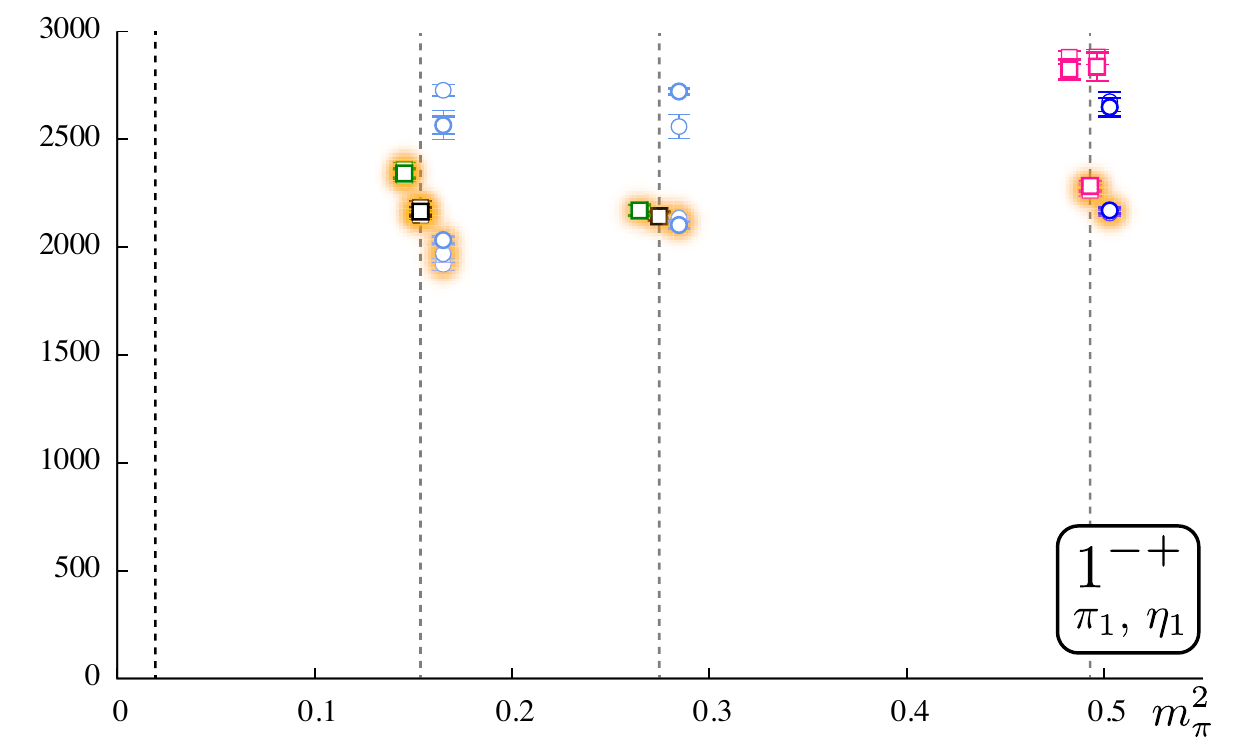} 
\includegraphics[width=.32\textwidth]{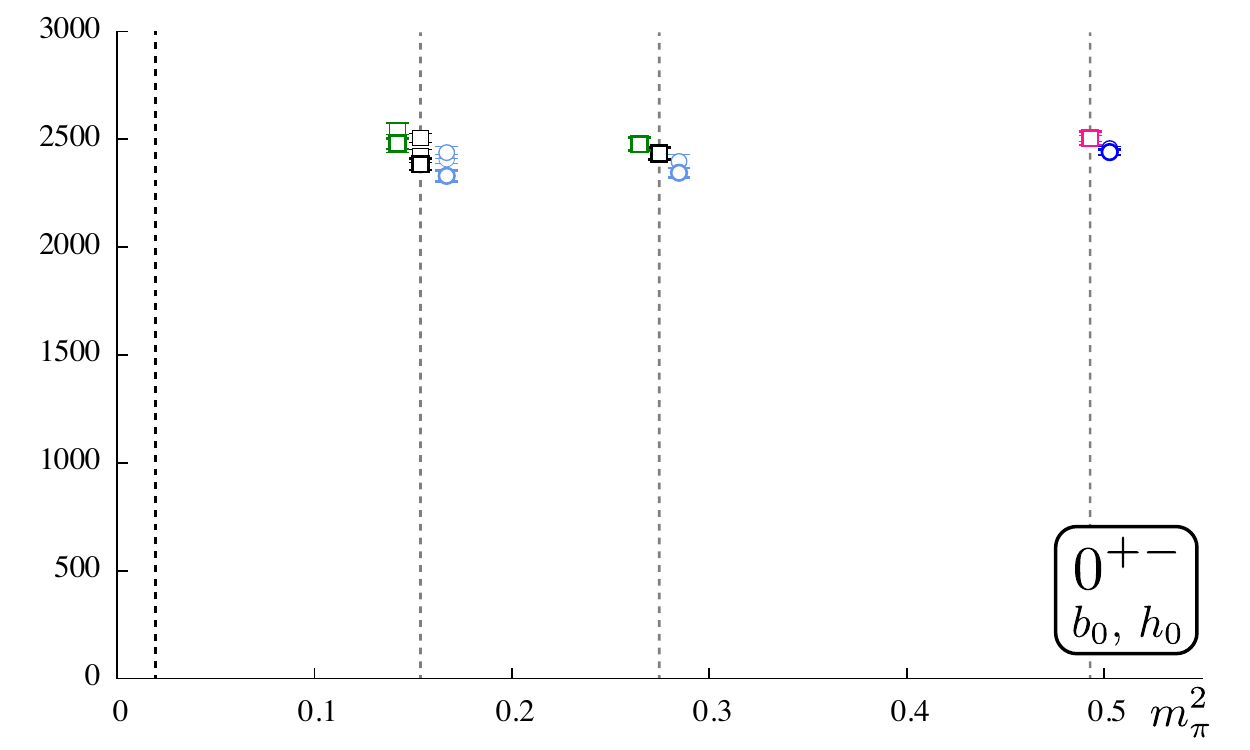} 
\includegraphics[width=.32\textwidth]{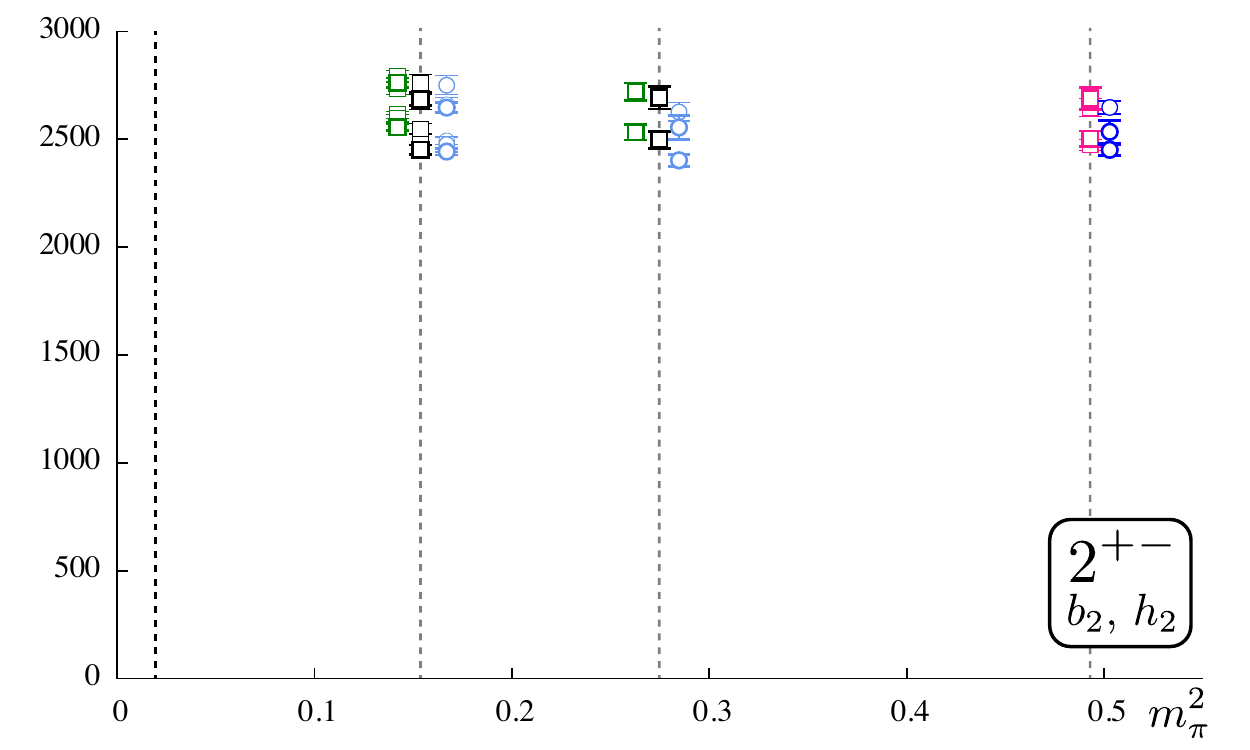} 
\caption{See caption of Figure \ref{0mp1mm}\label{Jmm}}
\end{figure*}

Figures \ref{0mp1mm}, \ref{Jpp} and \ref{Jmm} 
show the quark mass and volume dependence of the extracted isoscalar and isovector spectra. 

In general, the extracted spectrum is fairly consistent across quark masses. There are some cases, such as the second level in $3^{+-}$, that are not cleanly extracted at the lowest pion mass. 

We refrain from performing extrapolations of the masses to the limit of the physical quark masses, since, as we have already pointed out, we expect most excited states to be unstable resonances. A suitable quantity for extrapolation might be the complex resonance pole position, but we do not obtain this in our simple calculations using only ``single-hadron" operators. 

We discuss the specific case of the $0^{-+}$ and $1^{--}$ systems in the next subsections.

\subsection{The low-lying pseudoscalars: $\pi, \eta, \eta'$} \label{pieta}

\begin{figure}
 \centering
\includegraphics[width=.46\textwidth
]{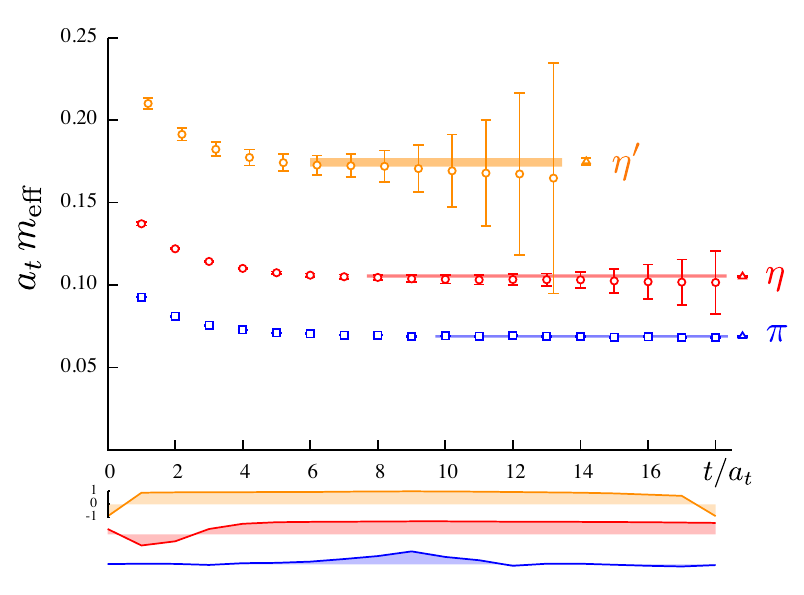} 
\caption{Effective masses (Eqn \ref{meff} with $\delta t = 3 a_t$) of principal correlators for the lowest level in the isovector sector ($\pi$, $t_0/a_t = 6$) and the lowest two levels in isoscalar sector ($\eta, \eta'$, $t_0 /a_t = 4$), on the $m_\pi = 391 \, \mathrm{MeV}$ lattice of volume $24^3\times 128$. Triangles at right show the mass determined with a two exponential fit to the principal correlator. The rectangles show a constant fit to the effective mass over the time-range indicated, in good agreement with the two exponential fits. Stacked graphs below the main graph display the timeslice correlation with respect to a reference time of $9 a_t$.\label{eta}}
\end{figure}

In lattice calculations of the type performed in this paper, where isospin is exact and electromagnetism does not feature, the $\pi$ and $\eta$ mesons are exactly stable and $\eta'$ is rendered stable since its isospin conserving $\eta \pi \pi $ decay mode is kinematically closed. Because of this, many of the caveats presented in Section \ref{interpret} do not apply. Figure \ref{eta} shows the quality of the principal correlators from which we extract the meson masses, in the form of an effective mass,
\begin{equation}
m_\mathrm{eff} = \frac{1}{\delta t} \log \frac{\lambda(t)}{\lambda(t+\delta t)}, \label{meff}
\end{equation}
for the lightest quark mass and largest volume considered. The effective masses clearly plateau and can be described at later times by a constant fit which gives a mass in agreement with the two exponential fits to the principal correlator that we typically use.

\begin{figure}
 \centering
\includegraphics[width=.49\textwidth
]{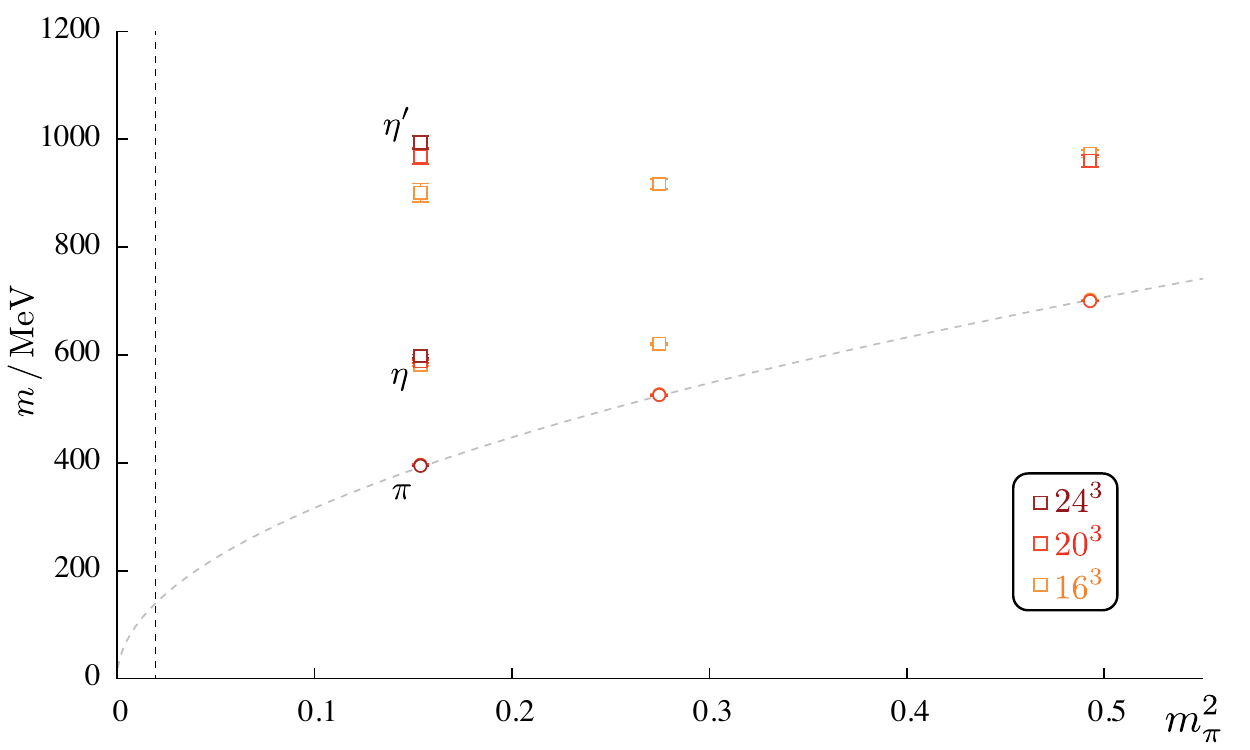} 
\caption{Quark mass and volume dependence of low-lying pseudoscalar mesons. Plotted against $m_\pi^2$ in units of $\mathrm{GeV}^2$. \label{pseudo}}
\end{figure}

Figure \ref{pseudo} indicates the detailed quark mass and volume dependence of the $\eta$ and $\eta'$ mesons. We have already commented on the unexplained sensitivity of the $\eta'$ mass to the spatial volume at $m_\pi = 391\,\mathrm{MeV}$, and we note that since only a $16^3$ volume was used at $m_\pi = 524\,\mathrm{MeV}$, the mass shown there may be an underestimate. 

\begin{figure}
 \centering
\includegraphics[width=.49\textwidth
]{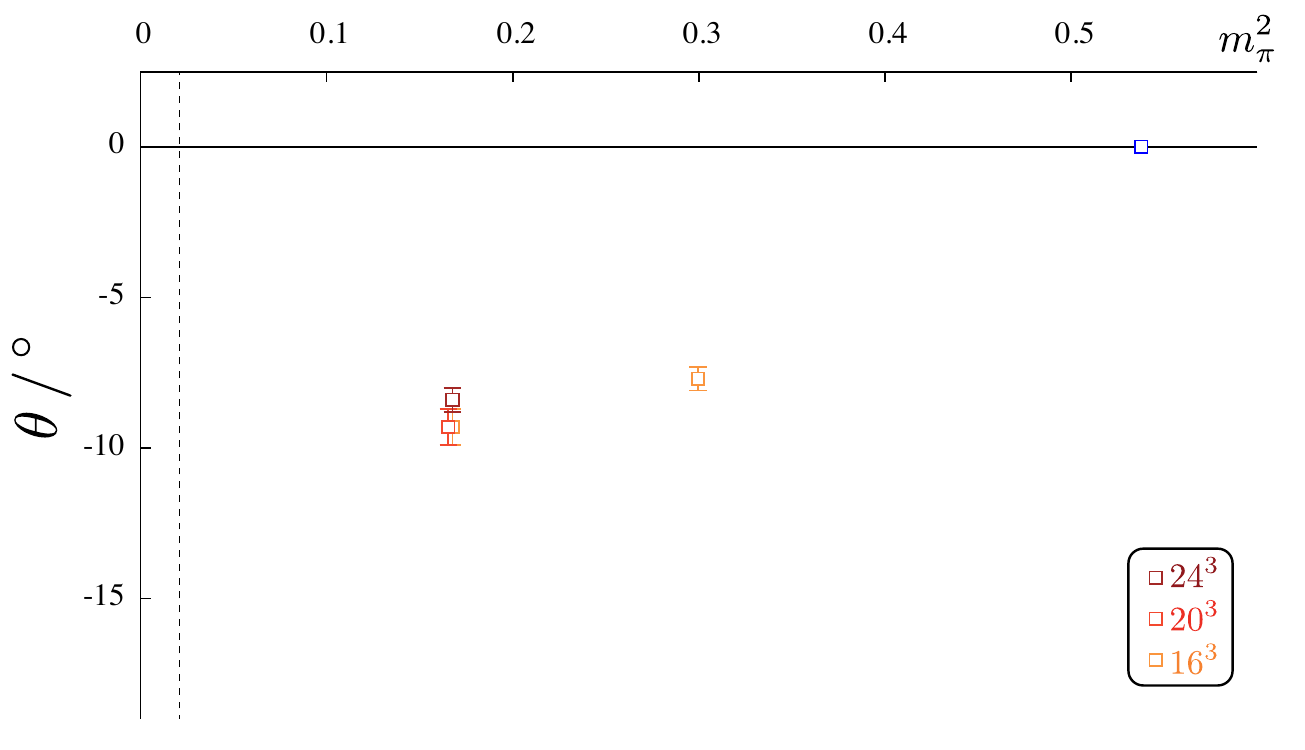} 
\caption{Octet-singlet mixing angle, $\theta = \alpha - 54.74^\circ$, for $\eta, \eta'$.  Plotted against $m_\pi^2$ in units of $\mathrm{GeV}^2$.\label{eta_mix}}
\end{figure}

Figure \ref{eta_mix} shows the octet-singlet basis mixing angle, $\theta = \alpha - 54.74^\circ$, which by definition must be zero at the $SU(3)_F$ point\footnote{
Here we are using a convention where $|\eta\rangle = \cos \theta |\mathbf{8}\rangle - \sin \theta |\mathbf{1}\rangle$, $|\eta'\rangle = \sin \theta |\mathbf{8}\rangle + \cos \theta |\mathbf{1}\rangle$ with $\mathbf{8}, \mathbf{1}$ having the sign conventions in Eqn \ref{SU3basis}.
}
. While we have no particularly well motivated form to describe the quark mass dependence, it is notable that the trend is for the data to approach a phenomenologically reasonable value $\sim -10^\circ$ \cite{Feldmann:1999uf, Escribano:2007cd, Thomas:2007uy, PDG}.  

\subsection{The low-lying vector mesons: $\rho, \omega, \phi$}

\begin{figure}
 \centering
\includegraphics[width=.46\textwidth
]{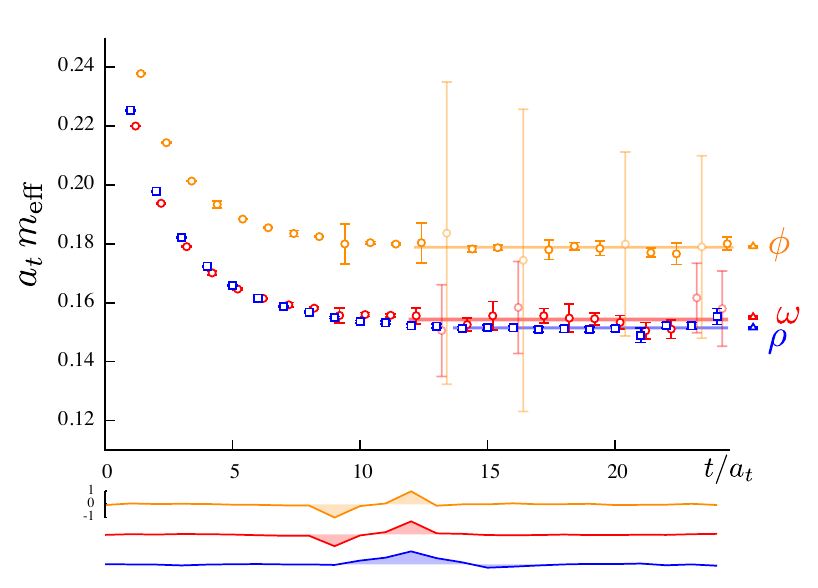} 
\caption{Effective masses (Eqn \ref{meff} with $\delta t = 3 a_t$) of principal correlators for the $T_1^{--}$ irrep. The lowest level in the isovector sector ($\rho$, $t_0/a_t = 9$) and the lowest two levels in isoscalar sector ($\omega, \phi$, $t_0 /a_t = 6$), on the $m_\pi = 391 \, \mathrm{MeV}$ lattice of volume $24^3\times 128$. 
Triangles at right show the mass determined with a two exponential fit to the principal correlator. The rectangles show a constant fit to the effective mass over the time-range indicated, in good agreement with the two exponential fits. Stacked graphs below the main graph display the timeslice correlation with respect to a reference time of $12 a_t$.\label{omega}}
\end{figure}

\begin{figure}
 \centering
\includegraphics[width=.49\textwidth
]{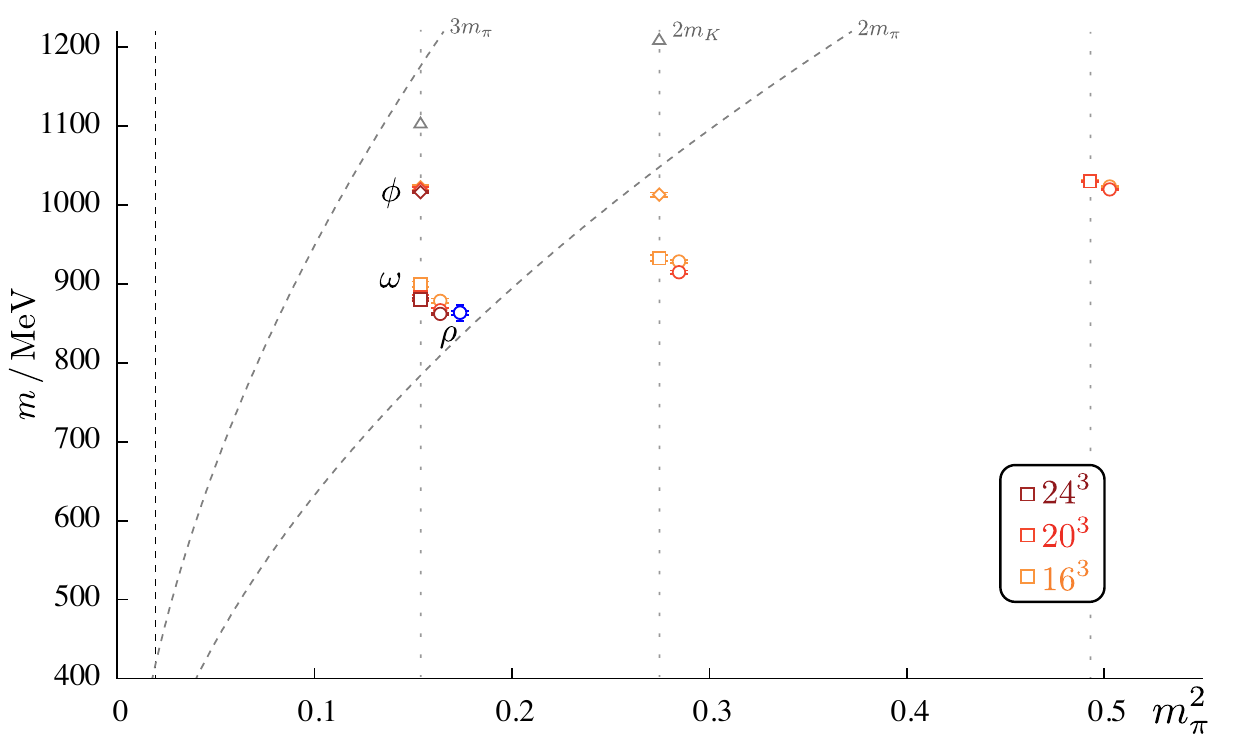} 
\caption{Quark mass and volume dependence of the low-lying vector mesons. $\rho$ masses (circles) shifted to the right for clarity (the vertical dashed lines show the actual pion mass). The blue circle (shifter further to the right) shows the $\rho$ as a resonance as determined in \protect\cite{Dudek:2012xn} - inner errorbars indicate the very small statistical uncertainty on the mass, outer bars show the hadronic width. Dashed lines show the decay threshold into two pions (relevant for the $\rho$) and three pions (relevant for the $\omega$). Triangles show the $K\overline{K}$ threshold relevant for decay of the $\phi$. Plotted against $m_\pi^2$ in units of $\mathrm{GeV}^2$ \label{vector}}
\end{figure}

Figure \ref{omega} shows the effective masses of $\omega, \phi$ and $\rho$ principal correlators on the $m_\pi = 391\,\mathrm{MeV}$, $24^3\times 128$ lattice. The splitting between the $\rho$ and $\omega$ is small but statistically significant, reflecting the small disconnected contribution at large times in this channel. At the pion masses presented in this paper, the $\omega$ and $\phi$ mesons are kinematically stable against decay into their lowest threshold channels, $\pi\pi\pi$ and $K\overline{K}$. In Figure \ref{vector} we show the quark mass and volume dependence of the low lying vector mesons along with the relevant threshold energies. The $\rho$ is unstable at our lightest pion mass and in the figure we show the $\rho$ ``mass" extracted from analysis featuring only fermion bilinear operators (as performed in this paper). We also show the $\rho$ as a resonance (in blue), determined in \cite{Dudek:2012xn} using a larger basis featuring also $\pi\pi$-like constructions and a L\"uscher analysis. We observe that the more naive calculation gives, on any given volume, a single level within the hadronic width of the resonance, as we have proposed earlier.

\begin{figure}
 \centering
\includegraphics[width=.49\textwidth
]{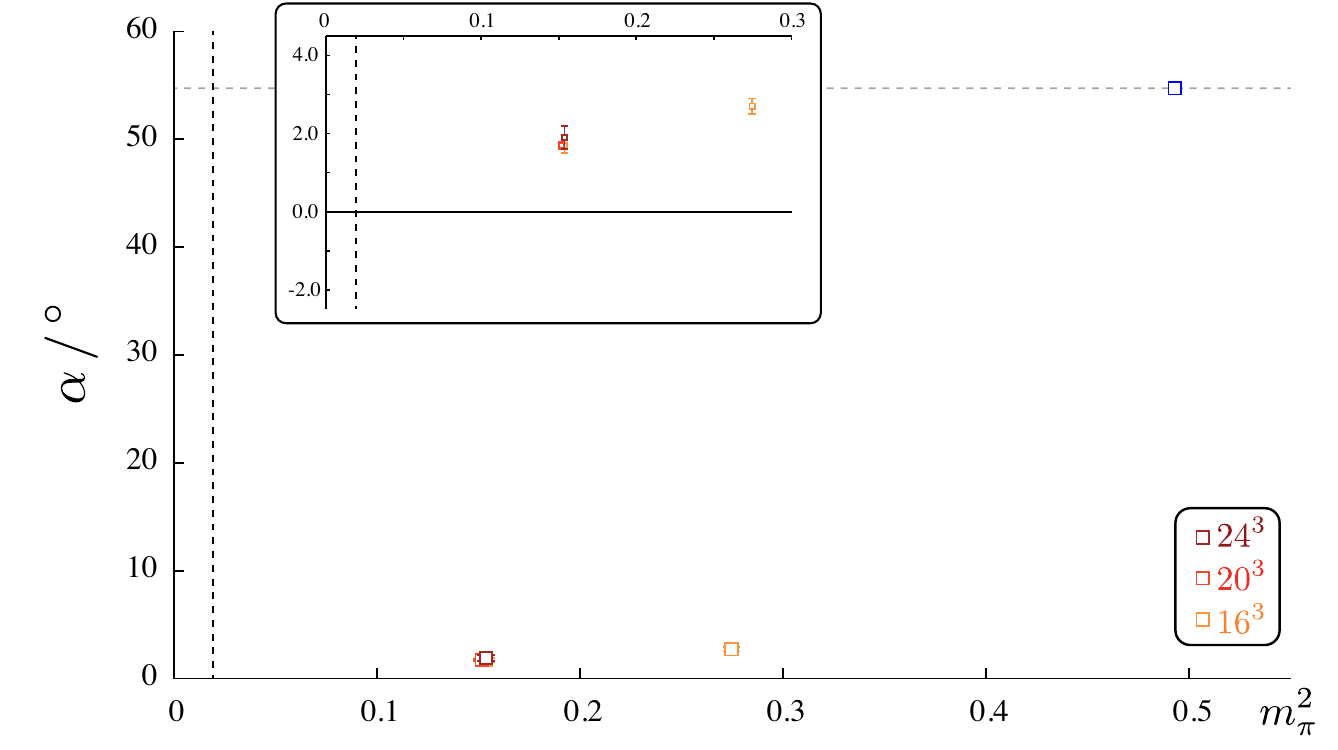} 
\caption{Flavor basis mixing angle $\alpha$ for $\omega, \phi$. Plotted against $m_\pi^2$ in units of $\mathrm{GeV}^2$. \label{omega_mix}}
\end{figure}

Figure \ref{omega_mix} shows the mixing angle in the flavor basis, $\alpha$, for the $\omega/\phi$ system. This angle must be $54.74^\circ$ at the $SU(3)_F$ point while it is \mbox{clearly very small for $m_\ell < m_s$ -- it is possible} for the angle to fall very rapidly if the disconnected contributions remain small, while the connected contributions $-\mathcal{C}^{\ell\ell}$, $-\mathcal{C}^{ss}$ start to differ significantly. The trend toward a very small mixing angle at the physical quark mass is in approximate agreement with the value $3.2(1)^\circ$ extracted from a model fit describing experimental vector to pseudoscalar radiative transitions \cite{Escribano:2007cd}.

\clearpage

\section{Phenomenology} \label{pheno}

A phenomenology of the isovector spectrum extracted on these lattices was presented in \cite{Dudek:2011bn, Dudek:2010wm}. Through observation of patterns of near degeneracy in the spectrum, and consideration of the relative values of extracted spectral overlaps $\big\langle \mathfrak{n} \big| \mathcal{O}^\dag \big| 0 \big\rangle$ for a set of operators $\mathcal{O}$ with fairly simple model interpretations, a description was provided that has the bulk of the spectrum being $q\bar{q}$ constructions featuring orbital angular momentum, as it is in constituent quark models. In addition a set of states was identified which did not fit into this picture; these were interpreted as being hybrid mesons, in which a $q\bar{q}$ pair in a color octet is coupled to a chromomagnetic gluonic excitation. The lightest such set was found with quantum numbers $(0,1,2)^{-+},\, 1^{--}$ where only the $1^{-+}$ member is manifestly exotic. 

We find that the isoscalar spectra presented here also generally fit well into this phenomenology -- since there are no dramatic changes in the gross structure of the spectrum as we vary the quark mass (apart from the required merging of two states into the octet at the $SU(3)_F$ point), we can use Figure \ref{840_V24} to illustrate our typical isoscalar spectrum.

We observe that there is very little light-strange mixing in most of the spectrum, and generally the $\tfrac{1}{\sqrt{2}}\left( |u\bar{u}\rangle + |d\bar{d}\rangle \right)$ dominated state is very close in mass to the corresponding isovector state, suggesting that the role of $q\bar{q}$ annihilation is relatively small. The $|s\bar{s}\rangle$ dominated states are generally a little heavier as we'd expect from the heavier strange quark mass. There appears to be a general trend that the splitting between the dominantly light and dominantly strange states reduces as we go higher in the spectrum, which might be interpreted as excitation energies for light quarks being slightly larger than for strange quarks.

These observations regarding the hidden light/strange make-up are in qualitative agreement with the standard phenomenology of some experimentally observed states, in particular the lightest $2^{++}$ mesons: $a_2(1320),\,f_2(1270),\, f'_2(1525)$. Relative rates of decay of $f_2, f_2'$ into $\pi\pi$ and $K\bar{K}$ final states, as well as the $\gamma\gamma$ widths \cite{PDG}, are compatible within models with only a very small amount of hidden light-strange mixing.

There are clear exceptions to the general observation that light-strange mixing is small, The case of the lowest $0^{-+}$ states, corresponding to the $\eta$ and $\eta'$, was already discussed in Section \ref{pieta}. They appear to have mixing much closer to the octet, singlet representations of $SU(3)_F$ respectively, in agreement with the conventional phenomenology of these states. The first excited state pair in the $0^{-+}$ channel, although not determined with high statistical precision, appears to have a much smaller mixing angle. The experimental situation regarding excited isoscalar pseudoscalars, $\eta(1295),\, \eta(1405), \eta(1475)\ldots$, is not sufficiently clear for us to make any firm comparisons.

\begin{figure}[h!]
 \centering
\includegraphics[width=.49\textwidth]{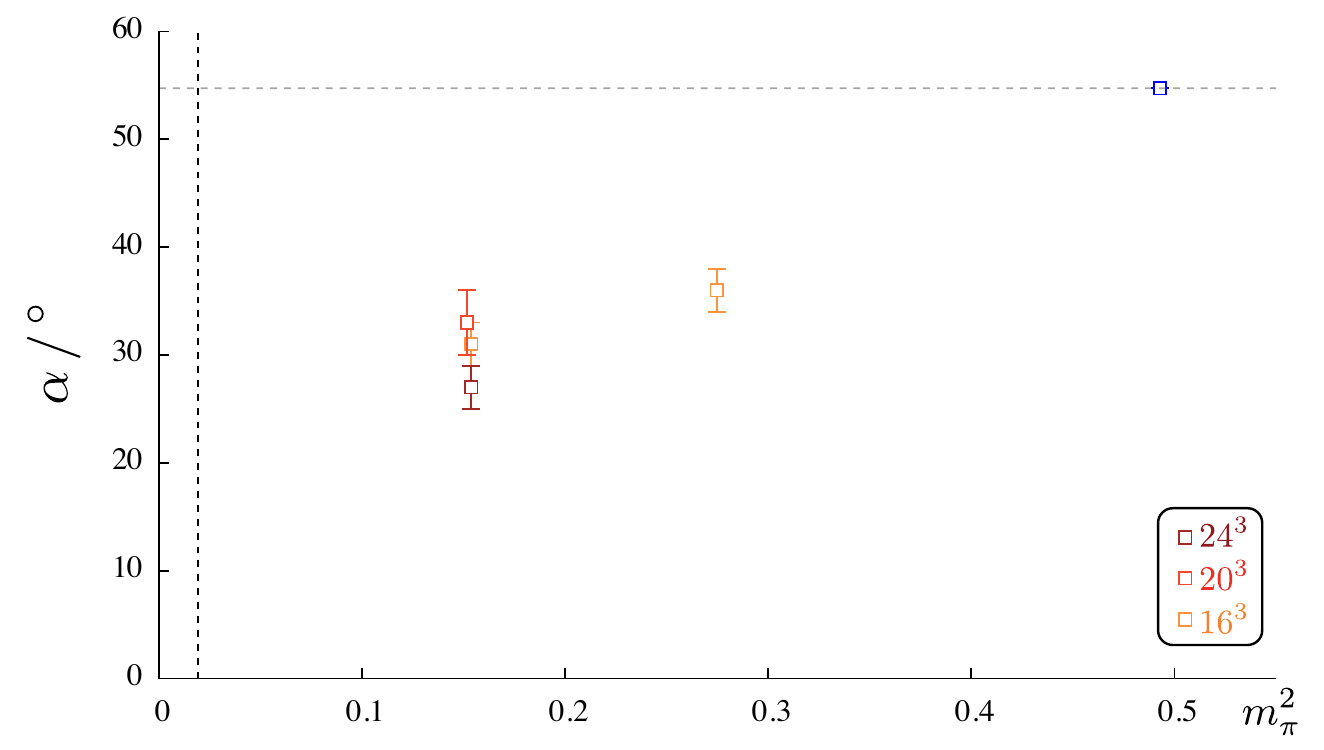} 
\caption{Flavor basis mixing angle $\alpha$ for lightest two $1^{++}$ states. Plotted against $m_\pi^2$ in units of $\mathrm{GeV}^2$. \label{f1_mix}}
\end{figure}

The other notable exception to the observation of ideal mass mixing is $1^{++}$, where both the lowest pair and the first set of excitations show significant mixing. The mixing angle for the lowest two states is shown in Figure \ref{f1_mix}, where we observe a significant departure from $\alpha \approx 0$. At the $SU(3)_F$ point there remains a large disconnected contribution splitting the singlet state from the octet. We can compare this pair with the experimental $f_1(1285), \, f_1(1420)$; these states are relatively narrow as they are forbidden to decay into a pair of pseudoscalars, the channel of largest phase-space, instead decaying into final states of higher multiplicity. The radiative decays of $f_1(1285)$ to $\gamma \rho$ and $\gamma \phi$ suggest a mixing angle $\alpha = 21(5)^\circ$ (following the formalism presented in \cite{Close:1997nm}, using current PDG averages \cite{PDG}), which is comparable to our extracted angles.

A clear set of isoscalar exotic $J^{PC}$ mesons is extracted. As seen in Figure \ref{840_V24}, there is a pair of isoscalars a little heavier than each of the isovector states with $1^{-+}, \, 0^{+-}$ and the two $2^{+-}$ states. In \cite{Dudek:2011bn} the distribution of $J^{PC}$ exotics observed in the isovector spectrum was proposed to be due to a chromomagnetic gluonic excitation coupling to a color-octet $q\bar{q}$ pair in an $S$-wave ($1^{-+}$) or in a $P$-wave ($0^{+-}, 2^{+-}, 2^{+-}$). The isoscalar spectrum has the same systematics. There is seen to be some degree of hidden light-strange mixing for the $1^{-+}$ and possibly the $0^{+-}$, but little or nothing for the $2^{+-}$ states. A large splitting between singlet and octet remains for the $1^{-+}$ at the $SU(3)_F$ point, suggesting that quark annihilation is important for these states. To date there are no experimental candidates for isoscalar exotic $J^{PC}$ mesons.

States with large overlap onto operators featuring a chromomagnetic construction are, as they were in the isovector spectrum, identified in the non-exotic spectrum at an energy scale comparable to the lightest $1^{-+}$ states. We argue that they make up the lightest hybrid supermultiplet with $(0,1,2)^{-+}, 1^{--}$, corresponding to a chromomagnetic excitation coupled to color octet $q\bar{q}$ in an $S$-wave. There are hints for some light-strange mixing in these states, and at the $SU(3)_F$ point these states are all compatible with having some octet-singlet splitting. Thus, it appears that annihilation contributions are a generic feature for the lightest hybrid supermultiplet, even in channels like $1^{--}$ where the other mesons, associated with $q\bar{q}$ structure, apparently feel very little from annihilation. Our model description of hybrid mesons is that they have structure $\big[q\bar{q}\big]_{\mathbf{8}_c} \otimes G_{\mathbf{8}_c}$, and since the quarks are in a color-octet, rather than the conventional color-singlet, their annihilation systematics need not be the same as in regular mesons. We were not able to resolve the non-exotic partners of the $0^{+-}, 2^{+-}, 2^{+-}$ exotics constructed from a chromomagnetic excitation coupled to color octet $q\bar{q}$ in an $P$-wave -- these states lie in an energy region where the positive parity spectrum is becoming dense and hard to disentangle, but interested readers should see Ref \cite{Liu:2012ze} where such states {\it were} successfully extracted in a comparable calculation of charmonium.

We remind the reader that we were unable to propagate correlators featuring glueball operators through our variational solution and thus are not able to discuss the role, if any, that glueball basis states play in the spectrum. In passing we note that the $2^{++}, 0^{-+}$ glueballs determined in Yang-Mills calculations \cite{Morningstar:1999rf} would be located at the very top of our extracted spectrum. The lightest glueball is expected to have scalar quantum numbers, but we have not presented results for the scalar sector as it is clear that two-hadron operator constructions will be essential for accurately determining the spectrum in this channel.

\section{Outlook}  \label{outlook}

We have taken advantage of favorable properties of the distillation framework to compute the disconnected diagrams required for isoscalar mesons with high statistical precision. Furthermore, the factorization between operator construction and quark propagation inherent in distillation allowed us to 
utilize a large basis of fermion bilinear operators of various spatial structures, including a selection featuring a chromomagnetic gluonic construction, from which matrices of correlation functions could be evaluated. Variational analysis of these lead to detailed excited state spectra that were interpreted phenomenologically. While the extracted spectra, which showed very little volume dependence, cannot be complete, as they clearly lack the dense and strongly volume-dependent spectrum of expected multi-hadron states, we have argued that they are a good guide to distribution of resonant meson excitations.

Ultimately comparison to experimental observations requires more than a semi-quantitative description of the position of various meson excitations -- calculations must observe resonances, and to do this we must resolve the complete spectrum of eigenstates in a finite volume. We observed in \cite{Dudek:2012xn} that to practically achieve this it is mandatory to include operators which resemble multiple-mesons with relative momentum. With the complete discrete spectrum in a given energy region determined, 
L\"uscher methods~\cite{Luscher:1990ux,Luscher:1990ck,Luscher:1991cf,Rummukainen:1995vs,Kim:2005gf,Christ:2005gi} and inelastic extensions~\cite{He:2005ey,Briceno:2012yi,Hansen:2012tf,Guo:2012hv} can be applied to determine scattering amplitudes that may have a resonant interpretation.

For excited hadron spectroscopy, calculations at the physical pion mass are not yet warranted -- with a 140 MeV pion, a large number of kinematic thresholds would be open for even the lowest resonances, and this would demand involved coupled-channel analyses featuring channels of high multiplicity ($\pi\pi\pi, \,\pi\pi\pi\pi \, \ldots$). These coupled-channel analyses are still in their infancy even for simpler cases involving only two-body scattering channels. A sensible approach, given this, is to develop the techniques for extracting hadron scattering amplitudes for the case of heavier pions where there will be a restricted number of open channels, and then, once such techniques are mature, to push down the quark masses to their physical values.

Using what we know empirically, we have cause to be optimistic that the problem may not be as complicated as it could be. It is likely that we will find in QCD calculations, as was found experimentally, that true high-multiplicity final states are not significantly directly populated in hadron resonance decays, with most decays going through intermediate two-body states featuring isobar resonances.

In addition to studying the excited state resonant spectrum through evaluation of scattering amplitudes in finite-volume, there are a number of more straightforward extensions of the technology presented in this paper to phenomenologically interesting quantities. Computation of two-point correlation functions using {\it unsmeared} fermion bilinear operators at the source and distillation smeared operators at the time-varying sink would allow access to decay constants and distribution amplitudes of isoscalar and isovector mesons. Extension to three-point correlation functions with a local vector current insertion between distillation smeared source and sink operators would yield electromagnetic form-factors and radiative transition matrix elements \cite{Dudek:2009kk}. 

The results presented in this paper represent important progress toward
a description of isoscalar mesons within QCD.  Furthermore, they
suggest that the calculations proposed above can realistically be
attempted and hence give the prospect of a greater theoretical
understanding of the isoscalar meson sector.


\begin{acknowledgments}
We thank our colleagues within the Hadron Spectrum Collaboration. Special thanks is given to B\'alint Jo\'o for help with the GPU code. {\tt Chroma}~\cite{Edwards:2004sx} and {\tt QUDA}~\cite{Clark:2009wm,Babich:2010mu} were used to perform this work on clusters at Jefferson Laboratory under the USQCD Initiative and the LQCD ARRA project. Gauge configurations were generated using resources awarded from the U.S. Department of Energy INCITE program at Oak Ridge National Lab, the NSF Teragrid at the Texas Advanced Computer Center and the Pittsburgh Supercomputer Center, as well as at Jefferson Lab. RGE, PG and JJD acknowledge support from U.S. Department of Energy contract DE-AC05-06OR23177, under which Jefferson Science Associates, LLC, manages and operates Jefferson Laboratory. JJD also acknowledges the support of the U.S. Department of Energy Early Career award contract DE-SC0006765.  
CET acknowledges support from a Marie Curie International Incoming Fellowship, 
PIIF-GA-2010-273320, within the 7th European Community Framework Programme.

\end{acknowledgments}

\begin{widetext}
\bibliography{isoscalar}
\end{widetext}

\end{document}